\documentclass[10pt,twoside]{article}
\usepackage{lmodern}
\usepackage{fancyhdr}
\usepackage{titlesec}
\usepackage{cite}
\usepackage{caption}    
\usepackage{appendix} 
\usepackage{pifont}
\usepackage{stmaryrd}
\usepackage{setspace}
\usepackage{indentfirst}
\usepackage{amsmath,amssymb,amscd,bbm,amsthm,mathrsfs,dsfont}
\usepackage{amsmath,amssymb,amscd,bbm,amsthm,mathrsfs,dsfont}
\usepackage{xcolor}
\usepackage{picins}
\usepackage{graphicx}
\usepackage{multirow}
\usepackage{wrapfig} 
\usepackage{CJKutf8}
\usepackage{tikz}
\usetikzlibrary{arrows.meta, positioning}
\usepackage{algorithm}
\usepackage{algpseudocode}
 \label{alg:AOA}
    \renewcommand{\algorithmicrequire}{\textbf{Input:}}
    \renewcommand{\algorithmicensure}{\textbf{Output:}}
\usepackage{amsthm}
\usepackage{caption}
\newtheoremstyle{indented}
  {\topsep}      
  {\topsep}      
  {\itshape}     
  {\parindent}   
  {\bfseries}    
  {.}            
  { }            
  {}             

\theoremstyle{indented}
\newtheorem{theorem}{Theorem}[section]    
\newtheorem{lemma}[theorem]{Lemma}        

\newtheorem{definition}[theorem]{Definition} 

\usepackage{setspace}
\usepackage{changepage}

\usepackage{CJKutf8} 
\usepackage{etoolbox}
\AtBeginEnvironment{proof}{\setlength{\leftskip}{2em}\setlength{\rightskip}{0pt}}

\titleformat{\section}{\centering\large\bfseries}{\arabic{section}}{1em}{}
\newboolean{first}
\setboolean{first}{true}

\usepackage{amssymb}
\usepackage{mathptmx}
\usepackage{amsmath}
\usepackage{amsthm}
\usepackage{chngcntr}
\usepackage{booktabs}
\numberwithin{theorem}{section} 

\usepackage[utf8]{inputenc}
\usepackage{subfigure}
\usepackage{epstopdf}

\makeatletter
\newenvironment{breakablealgorithm}
{
		\begin{center}
			\refstepcounter{algorithm}
			\hrule height.8pt depth0pt \kern2pt
			\renewcommand{\caption}[2][\relax]{
				{\raggedright\textbf{\ALG@name~\thealgorithm} ##2\par}%
				\ifx\relax##1\relax 
				\addcontentsline{loa}{algorithm}{\protect\numberline{\thealgorithm}##2}%
				\else 
				\addcontentsline{loa}{algorithm}{\protect\numberline{\thealgorithm}##1}%
				\fi
				\kern2pt\hrule\kern2pt
			}
		}{
		\kern2pt\hrule\relax
	\end{center}
}
\makeatother
\begin{document}
	\begin{CJK}{UTF8}{gbsn} 
		\setlength\abovedisplayskip{2pt}
		\setlength\abovedisplayshortskip{0pt}
		\setlength\belowdisplayskip{2pt}
		\setlength\belowdisplayshortskip{0pt}
		\title{\bf \Large Radio Labeling of Strong Prismatic Network With Star \author{  Liming Wang$^{1}$, Feng Li$^{1, \ast}$, Linlin Cui$^1$} \\ \date{}}
		 \maketitle
		{ \qquad1 College of Computer Science, Qinghai Normal University, Xi'ning, 810008, P.R.China}
        	\footnote{$\ast$\, Corresponding author}
\footnote{Feng\,\,Li\,\,(E-mail:\,li2006369@126.com),\,\,Liming\,\,Wang\,\,(E-mail:\,W07180403@163.com),\,\,Linlin Cui\,\,
(E-mail:\,lin7232023@ 163.com) }
\begin{center}
\begin{minipage}{155mm}

{\bf \small Abstract}.\hskip 2mm {\small
The rapid development of wireless communication has made efficient spectrum assignment a crucial factor in enhancing network performance. As a combinatorial optimization model for channel assignment, the radio labeling is recognized as an NP-hard problem. Therefore, converting the spectrum assignment problem into the radio labeling of graphs and studying the radio labeling of specific graph classes is of great significance. For $G$, a radio labeling $\varphi: V(G) \to \{0, 1, 2, \ldots\}$ is required to satisfy $|\varphi(u) - \varphi(v)| \geq \text{diam}(G) + 1 -d_G(u, v)$, where ${diam(G)}$ and $d_G(u, v)$ are diameter and distance between $u$ and $v$. For a radio labeling $\varphi$, its $\text{span}$ is defined as the largest integer assigned by $\varphi$ to the vertices of $G$; the radio labeling specifically denotes the labeling with the minimal span among possible radio labeling. The strong product is a crucial tool for constructing regular networks, and studying its radio labeling is necessary for the design of optimal channel assignment in wireless networks. Within this manuscript, we discuss the radio labeling of strong prismatic network with star, present the relevant theorems and examples, and propose a parallel algorithm to improve computational efficiency in large-scale network scenarios.}

\end{minipage}
\end{center}


\section{Introduction}

The problem of wireless network frequency assignment is an optimization problem that aims to allocate a limited frequency band to different transmitters optimally under the constraint of no interference, in order to minimize the total system bandwidth usage. The frequency assignment problem is usually regarded as a complex combinatorial optimization problem, aiming to maximize the utilization of the system bandwidth resources and minimize the frequency reuse distance, which is key to achieving reliable communication.

The solution to spectrum assignment effectively alleviate the conflict between limited frequency resources and the increasing user demand, attracting attention from the academic community and being the subject of in-depth research. In 1980, Hale \cite{1} was framed as an optimal vertex labeling problem on graphs, with vertices representing transmitters and edges denoting interference relationships. Vertex labeling corresponds to channel assignment, and the span of the labeling equals the channel span. Specifically, let $G$ stand for the graph expressed as a tuple $ G = (V(G), E(G)) $, where $V(G)$ denotes the set of vertices and $E(G)$ denotes the set of edges. For any two vertices $u$ , $v$ in $G$, the distance is denoted as $d_G(u, v) $ and the diameter refers to the maximum of the shortest the distance separating any pair of vertices, defined as $\mathit{diam}(G) = \mathit{max}\{d_G(u, v) \mid u, v \in V(G) \}$. 

The graph product is an important tool for studying radio labeling, and the concept was first introduced in reference \cite{2}. The product graph mainly includes the Cartesian product, the strong product, the direct product, and the lexicographic product. For more studies on product graph, refer to references \cite{3,4,5,6,7}. The study concentrates on assigning radio labeling to strong prismatic network with star structure. The strong product of star and cycle serves as the basis for the network's construction. For the convenience of the subsequent discussion, we first provide the definitions of two simple graphs.

\begin{definition}

A star $S_n$ is a particular kind of tree that features a single central vertex $(c)$ and several leaves $v_i$ ($i=1,2,\dots,n$). All leaves are adjacent only to this central vertex and non-adjacent to each other, and it is a simple graph, and its diameter is $\text{diam}(S_n) = 2$.
\end{definition}

\begin{definition}
A cycle of length $i$ ($i \geq 3$) constitutes an ordered list of vertices and edges 
$C = \{v_0\ e_1\ v_1\ e_2\ \dots\ v_{i-1}\ e_i\ v_i\}$,
satisfying the following conditions, the starting vertex coincides with the ending vertex, and the remaining vertices $v_1, v_2, \dots, v_{i-1}$ in the sequence are all distinct. A cycle is a path with coinciding starting and ending points. A cycle is usually denoted as $C_m$, where $m$ represents the count of vertices together with the count of edges, and its diameter is $\text{diam}(C_m) = \lceil m/2 \rceil$.
\end{definition}

\begin{figure}[!ht]
    \centering
    \includegraphics[width=0.75\linewidth]{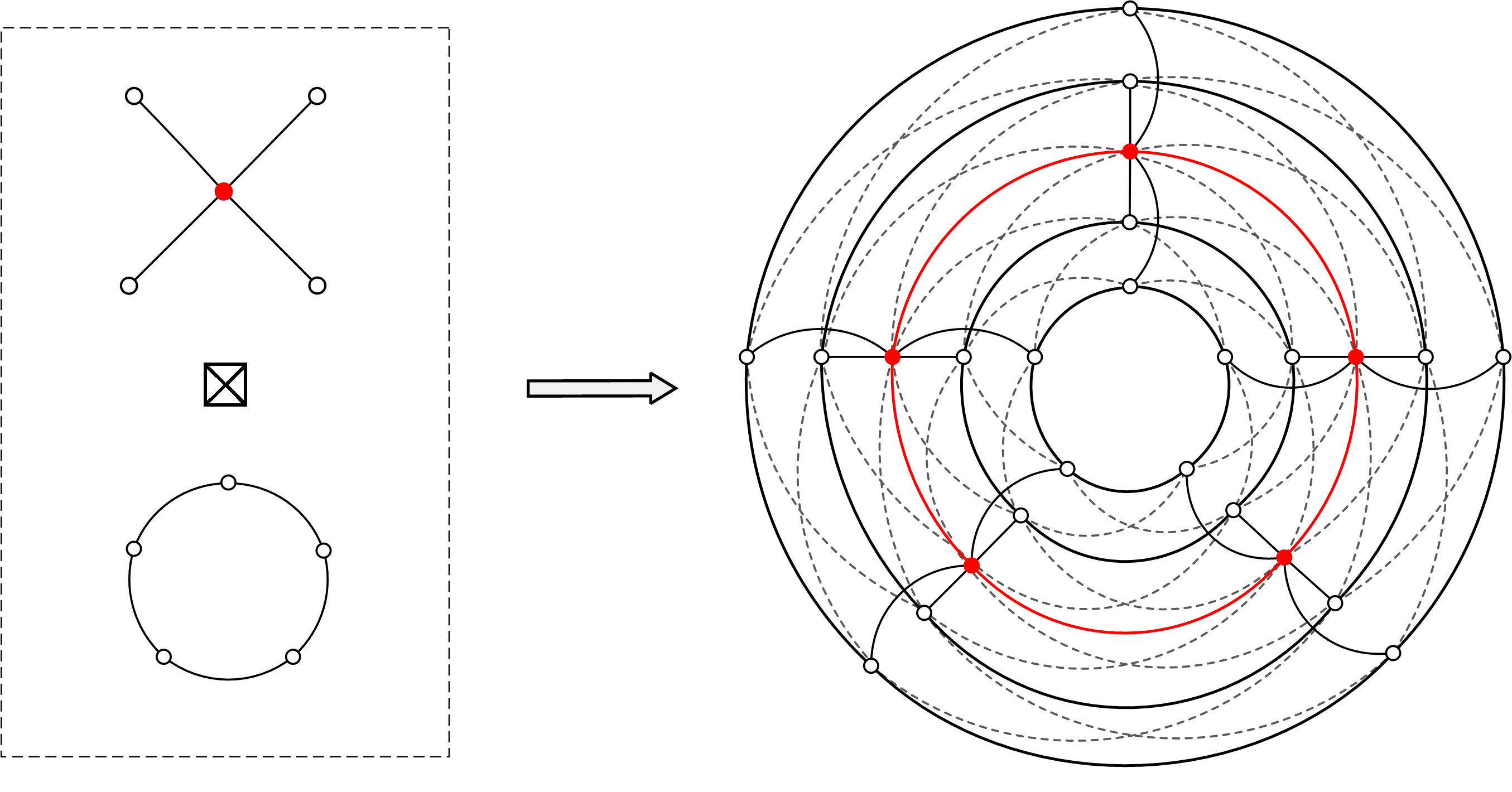}
    \caption{Strong prismatic network with 4-order stars.}
    \label{fig:placeholder1}
\end{figure}

The special graphs studied in this paper rely on strong product constructions. Specifically,

Let $G_1=(V_1,E_1)$ and $G_2=(V_2,E_2)$ be two undirected graphs. For any graphs $G_1 = (V_1, E_1)$ and $G_2 = (V_2, E_2)$, their strong product forms an undirected graph, commonly represented by $G = G_1 \boxtimes G_2$, with vertex set $V(G_1 \boxtimes G_2) = V_1 \times V_2$. Two separate vertices $u_1u_2$, $v_1v_2$ (where $u_1, v_1 \in V_1$ and $u_2, v_2 \in V_2$) are adjacent exactly when at least one of the following holds: (1) $u_1 = v_1$ and $(u_2, v_2) \in E_2$; (2) $u_2 = v_2$ and $(u_1, v_1) \in E_1$; (3) $(u_1, v_1) \in E_1$ and $(u_2, v_2) \in E_2$.

The strong prismatic network with star is composed of the strong product of star and cycle. Therefore, based on the definition of the strong product, we provide its definition.

\begin{definition}
The strong prismatic network with star is obtained by the strong product of star $S_n$ and cycle $C_m$, denoted as $G = S_n \boxtimes C_m$.
\end{definition}

With the aim of better grasping the defined network structure of the strong prismatic network with star, we have constructed the strong prismatic network with 4-order stars, as shown in Figure \ref{fig:placeholder1}.

\begin{lemma}[\cite{2}]
Let \(G_1\) and \(G_2\) be two connected undirected graph with $G=G_1\boxtimes G_2$ their strong product, then for any two separate vertices $u=(u_1,v_1)$, $v=(u_2,v_2)\in V(G)$, their distance in $G$ is:
\begin{equation}
d_G(u, v) = \mathit{max}\left\{d_{G_1}(u_1, u_2), d_{G_2}(v_1, v_2)\right\},
\end{equation}
where $d_{G_1}(u_1, u_2)$ denotes the distance between $u_1$ and $u_2$ in $G_1$, and $d_{G_2}(v_1, v_2)$ denotes the distance between $v_1$ and $v_2$ in $G_2$.
\end{lemma}

The diameter of strong product is the greatest distance between any pair of vertices, and its characteristics are governed by that distance.

\begin{lemma}[\cite{2}]
Let $G_1$ and $G_2$ be two simple connected undirected graphs, then the diameter of $G=G_1\boxtimes G_2$ is:
\begin{equation}
\mathit{diam}(G_1 \boxtimes G_2) = \mathit{max}\left\{\mathit{diam}(G_1),\ \mathit{diam}(G_2)\right\}.
\end{equation}
\end{lemma}

This article studies strong prismatic network with star constructed through strong product. As a composite graph structure, the strong product generates complex network topologies with higher regularity through the product operation of two basic graphs. These advantages have attracted attention from the academic community to the scheme of spectrum assignment in wireless networks based on product graphs. To characterize the channel assignment problem more precisely, the notion of radio labeling was first presented in \cite{8} put forward the idea of radio labeling. Radio labeling allocates a non-negative integer value to every individual vertex. To be more precise, radio labeling can be formally defined as.

\begin{definition}[\cite{8}]
The radio labeling of $G$ is a function $\varphi: V(G) \rightarrow \{0,1, 2, \ldots\}$ such that for any two distinct vertices $u, v \in V(G)$, satisfies the following inequality relation:
\begin{equation}
|\varphi(u) - \varphi(v)| \ge {diam}(G) + 1 -d_G(u, v),
\end{equation}

\noindent where ${diam}(G)$ denotes the diameter of $G$, $d(u,v)$ is the distance between vertices $u$ , $v$, and the span of a labeling $\varphi$ is defined as ${span}(\varphi) = \mathit{max}\{|\varphi(u) - \varphi(v)|,  u, v \in V(G)\}$. The radio labeling ${rn}(G)$ of a graph $G$ is:
\begin{equation}
\mathit{rn}(G) = \mathit{min}\bigl(\mathrm{span}(\varphi)\bigr),
\end{equation}
\end{definition}
\noindent where the minimum is obtained by considering all radio labelings of graph $G$. A radio labeling $\varphi$ of a graph $G$ is said to be optimal if its span satisfies ${span}(\varphi)$ = $ {rn}(G)$.

The \(L(2,1)\)-labeling concept, initially introduced by Griggs \cite{9} et al. Consider a graph $G$, an $L(2,1)$-labeling is then specified as a function $f: V(G) \rightarrow \{0,1,2,\ldots,k\}$ that assigns non-negative integers to vertices, satisfying:
\begin{equation}
|f(u) - f(v)| \ge
\begin{cases}
2, & \text{\textit{if }} d(u,v)=1;\\[2pt]
1, & \text{\textit{if }} d(u,v)=2.
\end{cases}
\end{equation}

Inspired by the $L(2,1)$-labeling, researchers extended the problem to $L(j,k)$-labeling. Chartrand \cite{10} et al. introduced a labeling scheme inspired by frequency modulation (FM) channel constraints to satisfy particular requirements based on distance requirements. In terms of radio labeling research, researchers made progress along several distinct directions, most of which are based on fundamental graph products and specific graph families. Path and cycle, because of their distinctive structural properties, can be combined with other graphs through product operations to generate more complex network topologies. These product graphs can serve as basic models for the study of complex graph structures. For instance, Liu and Zhu \cite{11} investigated the multilevel distance labeling problem. By analyzing the distance distribution among vertices in path and cycle, they determined the minimum radio labeling required for $G = P_n \square C_m $. Additionally, Kim \cite{12} et al. analyzed the radio labeling of $G = P_n \square K_m$, clarified the exact range of the radio number for this class of product graph. In reference \cite{13}, Niranjan et al. investigated the radio coloring of $G = K_n \square C_m$, and presented the exact radio labeling. Similarly, Wu \cite{14} et al. investigated the $L(j, k)$-labeling for $G = P_m \square C_n$. For different values of $j$, $k$, as well as the parity of $m$ and $n$, they precisely determined the corresponding labeling numbers. These results provide a theoretical foundation and optimization strategies for solving spectrum assignment problems in specific network topologies. Qi \cite{15} et al. investigated the $G = K_3 \boxtimes P_n$. By analyzing the distance relationships between vertices in the strong product, they determined its radio labeling. Meanwhile, Hong \cite{16} et al. determined the radio labeling of the graph formed by $C_4 \boxtimes P_n$, and constructed labeling scheme that achieves this optimum. In addition to path and cycle, trees are also common fundamental graphs for construct complex networks. Many studies combined trees with other fundamental graphs through products, achieving abundant results. Li \cite{17} et al. established exact formulas for the radio labeling of complete $m$-ary trees. Building on the preceding work, Tuzza \cite{18} studied radio labeling on hierarchical regular trees and obtained new results concerning lower bounds and their tightness. Shao \cite{19} et al. focused on the $L(2, 1)$-labeling problem of brick product graphs, proved that the $\lambda$-number of such graphs is either 5 or 6, and specified the conditions for the $\lambda$-number to be 5. Yeh \cite{20} summarized existing results on the labeling number of certain special graph classes under distance-2 conditions, analyzed the computational complexity of the problem, and provided bounds associated with the labeling numbers of various graph classes. This study \cite{21} provided an important reference for subsequent research in this field. This research examines the $2$-distance list coloring problem of planar graphs that possess a girth of $10$ or more. Simultaneously, proved such graphs are $(\Delta+2)$-list $2$-distance colorable when their maximum degree is $\Delta$. Li \cite{22} et al. investigated the optimal radio labeling for $ P(a, b) \square S_n$, and obtained favorable structural results. Sethuraman \cite{23} et al. investigated the radio labeling of biconvex split graphs by leveraging the unique structural properties of these graphs, precisely determined their radio labeling. In \cite{24}, the authors investigated the optimal radio labeling problem of $G = PN(n) \square P_m$, and provided the optimal radio labeling that achieve the smallest possible span for the graph.

Typically, deriving the radio labeling for general graphs poses considerable computational difficulty, and it is not possible to solve arbitrary instances exactly in polynomial time. Therefore, studying strong prismatic network with star is of significant importance. Through analyzing the parity of cycle, we establish tight bounds for the radio number of strong prismatic network with star and determine their exact values. Building on these bounds, we obtain optimal labeling schemes for even cycle and odd cycle. Furthermore, we develop an efficient parallel algorithm suitable for large-scale network and employ visualization models to clearly demonstrate the functional relationship between parameters and the optimal radio number.

In terms of organization, this work follows the structure below. Section 2 introduces the fundamental terminology and notation for star and cycle. Section 3 discusses the main results of this study. By analyzing the parity of cycle, we establish the upper and lower bounds for the radio labeling of strong prismatic network constructed through the strong product of star and cycle. Furthermore, this section also investigates the optimal radio labeling and illustrates the theoretical findings with concrete examples. Section 4 summarizes the content examined in this paper.

\section{Preliminaries}

All graphs analyzed in the present paper are simple, connected, undirected, and finite. This section presents fundamental concepts and known results that will be utilized throughout our analysis. It is particularly noteworthy that star, as a special class of tree structure, possesses extremely simplified connectivity. The regularity of this structure provides critical insights for investigating more complex graph.

Our research focuses on strong prismatic network with star. Graph products can combine elementary graph structures to form complex network topologies. The radio labeling for such product graph holds significant theoretical and practical importance, especially when addressing spectrum assignment challenges in multidimensional networks where both adjacency constraints and distance constraints must be satisfied.

In the process of exploring radio labeling tailored to product graphs, a great many related results have provided inspiration and ideas for this study, such as the following lemmas:

\begin{lemma}[\cite{13}]
Let $G = K_n \square C_m$, where $n$ is an even integer greater than 7. Then the radio number of $G$ can be expressed as:
\begin{equation}
rn(G) = 
\begin{cases} 
\dfrac{m^2n + 3mn - 2m + 10}{8}, & \textit{if } m \equiv 1 \pmod{4}; \\
\dfrac{m^2n + 5mn - 2m + 6}{8}, & \textit{if } m \equiv 3 \pmod{4}.
\end{cases}
\end{equation}
\end{lemma}

Lemma 1.1 provides the radio number for $G = K_n \square C_m$. We now consider a different class of graph, namely $G = P(a, b) \square S_n$.

\begin{lemma}[\cite{22}]
Consider $G = P(a,b)\square S_n$, the Cartesian product of $S_n$ and the rectangular mesh $P(a,b)$. The optimal radio labeling for $G$ is consequently:
\begin{equation}
rn(G) = 
\begin{cases} 
\dfrac{a^2b + 4ab^2n + 2ab^2 + 4abn + 4ab}{4}, & \text{if $a$ is even}; \\
\dfrac{2a^2bn + 2ab^2n - abn + a^2b + ab^2 + 5ab}{4}, & \text{if $a$ is odd}.
\end{cases}
\end{equation}

In the case where $a$ and $b$ are both odd, the radio labeling corresponding to $G$ is:
\begin{equation}
rn(G) = \frac{a^2b + 7b}{4} + \frac{ab^2 - b^2 + b^2n - 3bn + 3a}{2} + ab + ab^2n + 2abn - an + 7.
\end{equation}
\end{lemma}

These results reveal how the structural properties of component graphs influence the radio numbers of the product network, provided a reference basis for our research on strong prismatic network with star. Furthermore, this study also describes the influence mechanism of graph structural properties on radio labeling, offering important insights for the radio labeling problem of strong prismatic network with star constructed through the strong product, which is investigated in this paper. Building upon fundamental lemmas and the distinctive structural properties of star, we will obtain the main results of this manuscript through theoretical derivation, which completely characterize the optimal radio labeling of strong prismatic network with star.

Before presenting the main conclusions drawn from this paper, we first establish the necessary theoretical foundation. The parity of the cycle directly affects the topological structure of strong prismatic network with star. This topology determines the network radio labeling. Therefore, we provide a structural definition based on the parity of the cycle. The definitions required for even order cycle are first presented.

\begin{definition}
Let $G = S_n \boxtimes C_m$,  where $m \ge 4$ is even. The $G$ can be decomposed into $m$ sub-star $S_n(j)$ ($1 \le j \le m$), where $S_n(j) = \{(v_i, u_j) \mid 0 \le i \le n\}$. Each $S_n(j)$ is isomorphic to the star graph $S_n$, where: central vertex $c_j = (v_0, u_j)$ and leaves $v_j(k) = (v_k, u_j)$.
\end{definition}

\begin{definition}
For any $j \in [1, m-1]$, the edges between substars $S_n(j)$ and $S_n(j+1)$ satisfy: for all $(v_i, u_j) \in V(S_n(j))$ and $(v_l, u_{j+1}) \in V(S_n(j+1))$, $(v_i, u_j)(v_l, u_{j+1}) \in E(G)$ if and only if $v_i v_l \in E(S_n)$ or $v_i = v_l$ in $S_n$.
\end{definition}

\begin{definition} 
For any $i \in [1, m]$, let $j = i + m/2 \pmod{m}$. The subgraph induced by the vertex set $V(S_n(i)) \cup V(S_n(j))$ in $G$ is denoted by $G(i) \subseteq G$. Then $G$ contains $m/2$ pairwise disjoint such subgraphs $G(i)$. By the diameter property of the strong product, $\operatorname{diam}(G(i)) = m/2$.
\end{definition}


\begin{definition}
Let $ G = S_n \boxtimes C_m $, where $ m \geq 5 $ is odd. Define the critical path class $ P'(t) \subset G $ ($ t \in [1, n] $), where each path $P$ is of the form: $ P'(t) = v_a(1) \xrightarrow{\alpha} v_b\left(\frac{m+1}{2}\right) \xrightarrow{\beta} v_c(m) $, where $ v_a(1) \in S_n(1) $, $ v_b\left(\frac{m+1}{2}\right) \in S_n\left(\frac{m+1}{2}\right) $, $ v_c(m) \in S_n(m) $, and $ $a$ \neq $b$ \neq $c$ $; $\alpha$ denotes the distance between $v_a(1)$ and $v_b\left(\frac{m+1}{2}\right)$, $\beta$ denotes the distance between $v_b\left(\frac{m+1}{2}\right)$ and $v_c(m)$.
\end{definition} 
Based on vertex index relations, we pay special attention to the following three types of paths:
\begin{equation}
\begin{aligned}
P_1'(t) &= \{
(v_1,u_1)\xrightarrow{\frac{m+1}{2}}(v_3,u_{\frac{m+1}{2}})\xrightarrow{\frac{m+3}{2}}(v_2,u_m),\,
(v_3,u_1)\xrightarrow{\frac{m+3}{2}}(v_2,u_{\frac{m+1}{2}})\xrightarrow{\frac{m+1}{2}}(v_1,u_m)\}; \\
P_2'(t) &= \{
(v_2,u_1)\xrightarrow{\frac{m+1}{2}}(v_0,u_{\frac{m+1}{2}})\xrightarrow{\frac{m+3}{2}}(v_3,u_m)
\}; \\
P_3'(t) &= \{
(v_a,u_1)\xrightarrow{\frac{m+3}{2}}(v_b,u_{\frac{m+1}{2}})\xrightarrow{\frac{m+3}{2}}(v_c,u_m)
\mid 4\leq a,b,c\leq n
\}.
\end{aligned}
\end{equation}

Therefore, the following conclusion can be obtained:
\begin{equation}
\varphi(v) = \begin{cases}
\dfrac{m+3}{2}, & \text{\textit{if }} v \in \{(v_2, u_{\frac{m+1}{2}}), (v_3, u_{\frac{m+1}{2}})\}; \\
\dfrac{m+1}{2}, & \text{\textit{if }} v \in \{(v_1, u_{\frac{m+1}{2}})\}; \\
\dfrac{m-1}{2}, & \text{\textit{otherwise}}.
\end{cases}
\end{equation}

\begin{figure}[!ht]
    \centering
    \includegraphics[width=0.85\linewidth]{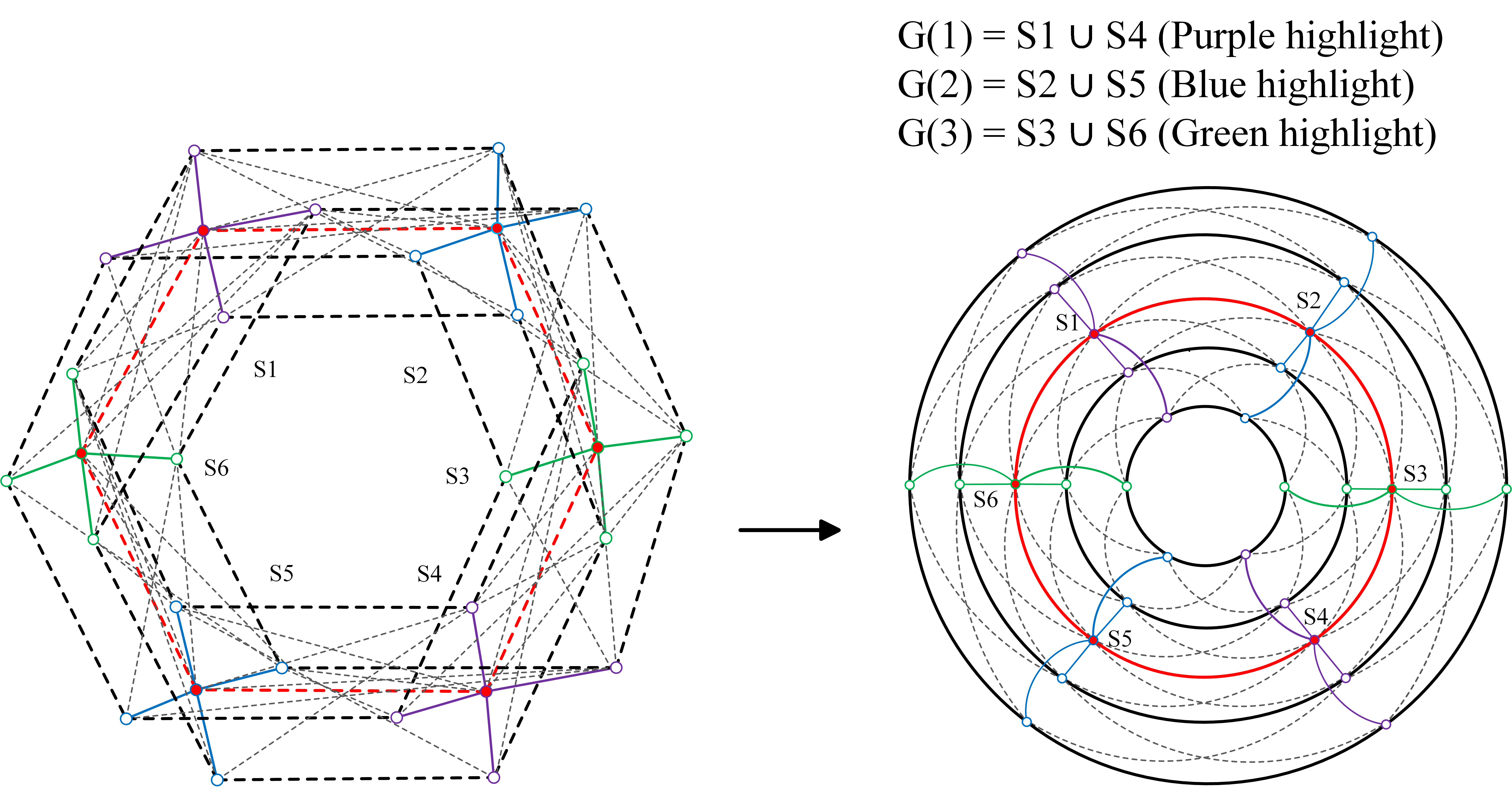}
    \caption{Substar decomposition.}
    \vspace{-3mm}
    \label{fig:placeholder2}
\end{figure}
\begin{definition}
Based on Definition 2.6, we partition $G$ as follows. Consists of vertices involved in critical paths, $ G' = S_n(1) \cup S_n(\frac{m+1}{2}) \cup S_n(m) $, Let $G'' = G \setminus V(G')$. For $i = 2, 3, \ldots, \frac{m-1}{2}$, define $j = i + \frac{m+1}{2} \pmod{m}$. The subgraph induced by the vertex set $v(S_n(i)) \cup v(S_n(j))$ is denoted by $G''(i) \subseteq G''$. Then $G''$ consists of $\frac{m-3}{2}$ pairwise disjoint $G''(i)$, and $\operatorname{diam}(G''(i)) = \frac{m+1}{2}$.
For any $ x \in S_n(t) $ and $ y \in S_n\left(t + \frac{m-1}{2}\right) $ ($ t \in [2, \frac{m-1}{2}] $), their distance is:
\begin{equation}
d(x_i, y_j) = \begin{cases}
\dfrac{m-1}{2}, & \text{if either } x \text{ or } y \text{ is a central vertex}; \\
\dfrac{m+1}{2}, & \text{if both } x \text{ and } y \text{ are leaves}.
\end{cases}
\end{equation}
\end{definition}

Note: If at least one vertex is central, a shorter path exists. If both are leaves, the path must pass through a central vertex, increasing the distance by 1. Thus, the distance is $\frac{m+1}{2}.$

To facilitate a better understanding of the definition of sub-star graphs in the definition, we provide illustrative examples in Figure \ref{fig:placeholder2}.

\begin{table}[htbp]
\centering
\caption{Relationship between vertex label differences and distances}
\label{tab:label_difference}
\begin{tabular}{c c c c}
\toprule
Condition & $\text{diam}(G) + 1 - d(u, v)$ & Value of $d(u, v)$ & Actual value of $|f(u) - f(v)|$ \\
& $= \dfrac{m+1}{2} - d(u, v)$ & & \\
\midrule
C1 & $1$ & $\dfrac{m-1}{2}$  & $1$ \\
C2 & $2$ & $\dfrac{m-3}{2}$ & $\geq 2$ \\
C3 & $\dfrac{m+1}{2} - k$ & $k$, where $1 \leq k < \dfrac{m-3}{2}$ & $\geq \dfrac{m+1}{2} - k$ \\
C4 & $\dfrac{m-1}{2}$ & $1$ (adjacent vertices) & $\dfrac{m-1}{2}$ or larger \\
\bottomrule
\end{tabular}
\end{table}
Analysis of radio labeling constraints for $G = S_n \boxtimes C_m$, where $m \geq 5$ is odd. For every possible graph distance $d(u, v)$, the table lists: (i) the minimum label difference required by the radio labeling condition $|f(u) - f(v)| \geq \operatorname{diam}(G) + 1 - d(u, v)$; (ii) the actual label difference achieved by our construction; and (iii) verification that the constraint is satisfied. This rigorously proves the tightness of the bound on the span.

After introducing the fundamental preliminaries, notational conventions, and relevant definitions, Section 3 of this paper will focus on proving the main theorems. First, through structural decomposition, the complex network is decomposed into analyzable substructural units. Subsequently, leveraging the distance properties of the strong product, the labeling constraints between vertices are established. Finally, based on the parity of the cycle, a complete optimal radio labeling scheme for strong prismatic network with star is proposed, while rigorously proving the consistency between the upper and lower bounds of the labeling numbers.

\section{Main Results}

This section investigates the radio labeling of strong prismatic network with star, with a focus on the parity of the cycle length $m$: Subsections 3.1 and 3.2 establish upper and lower bounds for even and odd $m$, respectively; Subsection 3.3 determines the exact radio number and presents an optimal labeling scheme.
\begin{figure}[ht]
    \centering
    \includegraphics[width=0.75\linewidth]{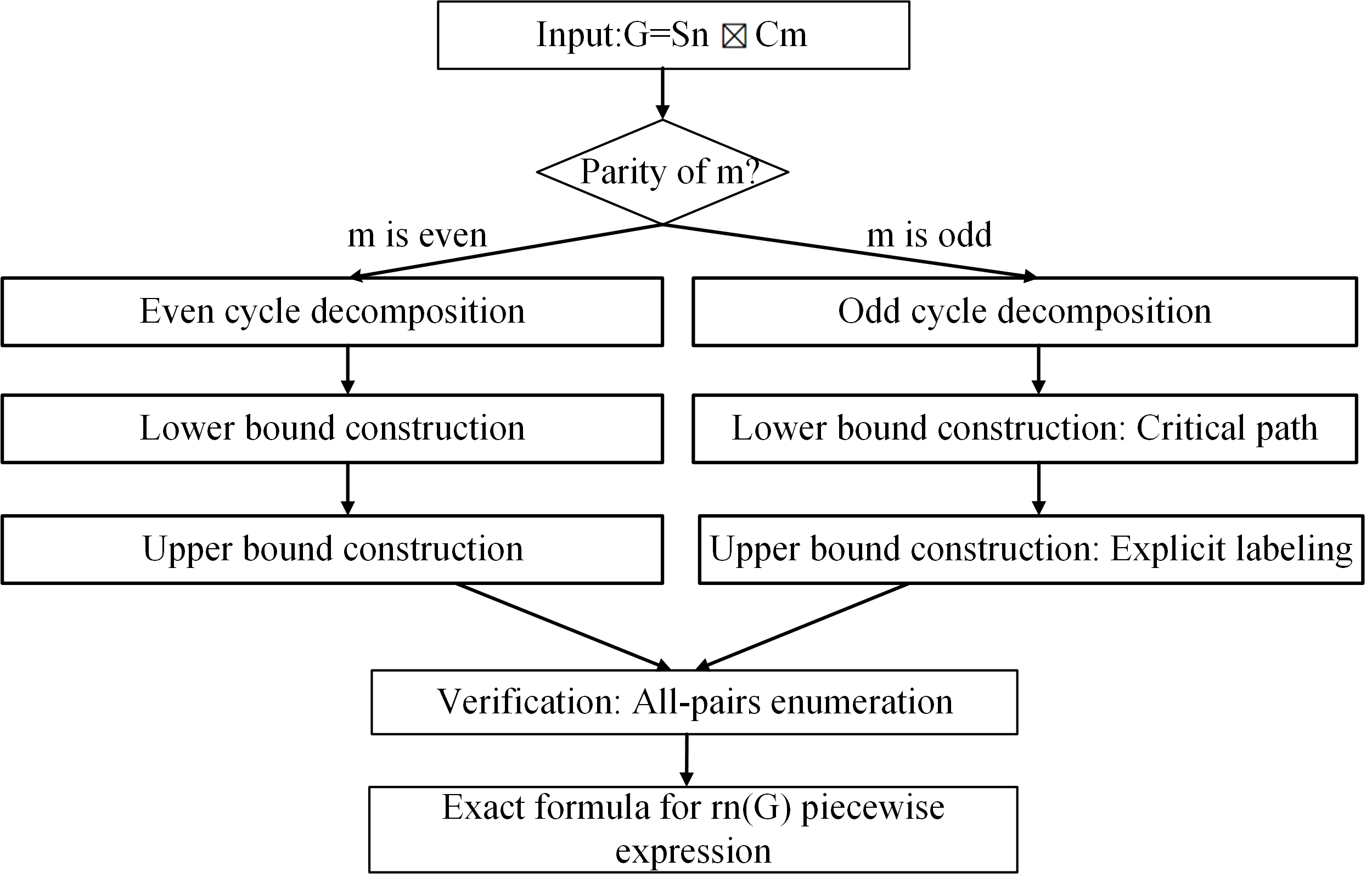}
    \caption{Substar decomposition.}
    \vspace{-3mm}
    \label{fig:placeholder3}
\end{figure}

In this paper, focusing on the strong prismatic network with star $G = S_n \boxtimes C_m$, we establish the tight bounds and optimal labeling scheme for the radio number $rn(G)$ by analyzing the parity of the cycle length $m$:
\begin{equation*}
rn(G)=
\begin{cases}
\dfrac{m^{2}+m(m-1)+(n-1)(m-2)}{2}, & \text{if $m$ is even and $m\ge 4$},\\[8pt]
\dfrac{m^{2}+4mn+5m-8n+18}{2},      & \text{if $m$ is odd and $m\ge 5$}.
\end{cases}
\end{equation*}

To systematically organize the research approach of this subsection, we have created the flowchart shown below, visually presenting the overall framework and key steps, thereby facilitating a clear understanding of the logical flow and research focus of this subsection.

\subsection{Bounds for the radio labeling of the strong prismatic network with star contain even cycle}

Based on the vertex partitioning structure defined at the end of Section 2, we examine the topological characteristics of strong prismatic network with star, thereby laying a crucial analytical foundation for proving the upper and lower bounds of its radio labeling when the cycle is of even order.

\begin{theorem}
Let $G = S_n \boxtimes C_m$ be the strong prismatic network with star, then the radio labeling of $G$ is:
\begin{equation}
rn(G) \ge \frac{m^2 + m(m-1) + (n-1)(m-2)}{2}.
\end{equation}
\end{theorem}

\begin{proof}
By Definition 2.3, let $c_j = (v_0, u_j)$ ($j = 1, \dots, m$) denote the central vertex of $S_n(j)$. For adjacent substars $S_n(j)$ and $S_n(j+1)$, the strong product implies $d_G(v_j, v_{j+1}) = 1$. Set $\varphi(c_1) = 0$.

By the definition of radio labeling, for any two distinct vertices $u, v \in V(G)$, we have:
\begin{equation}
|\varphi(u) - \varphi(v)| \ge {diam}(G) + 1 -d_G(u, v), \quad \forall u, v \in V(G)
\end{equation}

By Lemma 1.4, the diameter of $G$ is ${diam}(G) = \mathit{max}\{{diam}(S_n), {diam}(C_m)\}$. We further have ${diam}(G) = \frac{m}{2}$.

Based on the distance property of $G$ and the definition of the radio labeling, the radio labeling of $v_2$ is:
\begin{equation}
\varphi(v_2) \ge 0 + \frac{m}{2} + 1 - 1 = \frac{m}{2}.
\end{equation}

Assume $\varphi(v_{3}) \ge \varphi(v_2)$, by the distance property of $G$ and the definition of the radio labeling, the radio labeling of $v_3$ is:
\begin{equation}
\varphi(v_3) \ge \frac{m}{2} + \frac{m}{2} + 1 - 1 = m.
\end{equation}

By iterating the above formula for $j = 1, 2, \dots, m-1$, we obtain the radio labeling of $v_m$:
\begin{equation}
\varphi(v_m) \ge (m-1) \times \frac{m}{2}.
\end{equation}

For any leaf $v_j(k) = (v_k, u_j)$ ($k = 1, \dots, n$) in $S_n(j)$, the star structure ensures $d_G(v_j(k), c_j) = 1$. Thus, by the radio labeling condition, we obtain:
\begin{equation}
\begin{split}
\varphi(v_j(k)) &\ge \varphi(c_j) + \frac{m}{2} + 1 - 1 \\
&= \varphi(c_j) + \frac{m}{2}.
\end{split}
\end{equation}

For distinct leaves $v_j(k)$ and $v_j(l)$ in $S_n(j)$ satisfying $\varphi(v_j(k)) \ge \varphi(v_j(l))$, we have $d_G(v_j(k), v_j(l)) = 2$. Then from Equation(13) it follows that:
\begin{equation}
\begin{split}
|\varphi(v_j(k)) - \varphi(v_j(l))| &\ge \frac{m}{2} + 1 - 2 \\
&= \frac{m-2}{2}.
\end{split}
\end{equation}

Let $v_m(n)$ be the leaf of $S_n(m)$ with the largest radio labeling. The lower bound is satisfied as follows:
\begin{equation}
\varphi(v_m(n)) \ge \varphi(v_m) + \frac{m}{2} + (n-1) \times \frac{m-2}{2}.
\end{equation}
\

Substituting into inequality $(16)$,
\begin{equation}
\begin{aligned}
\varphi(v_m(n)) &\ge (m-1) \times \frac{m}{2} + \frac{m}{2} + (n-1) \times \frac{m-2}{2} \\
&= \frac{m^2 + m(m-1) + (n-1)(m-2)}{2}.
\end{aligned}
\end{equation}

Since $rn(G)$ is the minimum span of all radio labeling. Therefore, the lower bound of the radio labeling for $G$ is:
\begin{equation}
rn(G) \ge \frac{m^2 + m(m-1) + (n-1)(m-2)}{2}.
\end{equation}

This completes the proof.
\end{proof}

This establishes the lower bound for the radio number of the strong prismatic star network with even cycle order. The matching upper bound is given in Theorem 3.2.

\begin{theorem}
Let $G = S_n \boxtimes C_m$ be the strong prismatic network with star, then the radio labeling of $G$ is:
\begin{equation}
rn(G) \le \frac{m^2 + m(m-1) + (n-1)(m-2)}{2}.
\end{equation}
\end{theorem}

\begin{proof}
Define the central vertex $ c_j = (v_0, u_j) $ ($ j = 1, 2, \ldots, m $) of $ S_n(j) $. From Equation (13), the radio labeling of the vertex $v_j$ is:
\begin{equation}
\varphi(v_j) = (j-1) \times \frac{m}{2}.
\end{equation}

Let $c_j$ and $c_{j+1}$ be the central vertices of the two adjacent sub-star. Their label difference is:
\begin{equation}
\vert \varphi(v_j) - \varphi(v_{j+1}) \rvert = \frac{m}{2}. 
\end{equation}

For two non-adjacent central vertices $c_j$ and $c_{j+t}$, their distance in the network topology is $d(c_j, c_{j+t}) = t$. From the radio labeling condition, we obtain:
\begin{equation}
\vert \varphi(u) - \varphi(v) \rvert = \frac{m t}{2}. 
\end{equation}

According to the definition of the radio labeling and Equation (13), two non-adjacent central vertices should satisfy the following:
\begin{equation}
\begin{aligned}
\lvert \varphi(u) - \varphi(v) \rvert &\ge {diam}(G) + 1 - d(c_j, v_{c+t}) \\
&= \frac{m+2}{2} - t.
\end{aligned}
\end{equation}

For the $k$-th leaf $ v_j(k) = (v_k, u_j) $ of the sub-star $ S_n(j) $, its label is defined by Equation (16) as:
\begin{equation}
\varphi(v_j(k)) = \varphi(c_j) + \frac{m}{2} + (k-1) \times \frac{m-2}{2}.
\end{equation}

This is the radio labeling condition.

Let $v_j(k)$ and $v_j(l)$ be two leaves in the same star $S_n(j)$. By the star structure, their distance is $d_G(v_j(k), v_j(l)) = 2$. According to $|\varphi(u) - \varphi(v)| \ge {diam}(G) + 1 -d_G(u, v)$, the difference between the labels of $ v_j(k) $ and $ v_j(l) $ is:
\begin{equation}
\varphi(v_j(l)) - \varphi(v_j(k)) = (l-k) \times \frac{m-2}{2}.
\end{equation}

Consider two leaves $v_j(k) \in S_n(j)$ and $v_{j+t}(l) \in S_n(j+t)$ belonging to distinct stars. Their distance satisfies $d_G(v_j(k), v_{j+t}(l)) = t$. Consequently,

\begin{equation}
\begin{aligned}
\varphi(v_{j+t}(l)) - \varphi(v_j(k)) &= \frac{m}{2} + 1 - t \\
&= \frac{m t}{2} + (l-k) \times \frac{m-2}{2}.
\end{aligned}
\end{equation}

To determine the maximum label in $G$, set $\varphi(v_1) = 0$. The largest label occurs at the last leaf $v_m(n)$ of the final substar $S_n(m)$. Substituting $j = m$ and $k = n$ into $(13)$ gives:
\begin{equation}
\begin{aligned}
\varphi(v_m(n)) &= \varphi(v_m) + \frac{m}{2} + (n-1) \times \frac{m-2}{2} \\
&= \frac{m^2 + m(m-1) + (n-1)(m-2)}{2}.
\end{aligned}
\end{equation}

This completes the proof of the theorem.

Therefore, the upper bound of the radio number of $G$ is: 
\begin{equation}
rn(G) \le \frac{m^{2}+m(m-1)+(n-1)(m-2)}{2}.
\end{equation}

Which completes the proof of the theorem.
\end{proof}

Thus, we have established the tight bounds for the radio number of the strong prismatic network with star for even values of $m$. This lays the groundwork for the analysis of the odd case, the corresponding tight bounds and complete proofs for odd $m$ are provided in Section $3.2$.

\subsection{Bounds for the radio labeling of the strong prismatic network with star contain odd cycle}

Using the path structures and distance properties from Definitions 2.6–2.7, together with the radio labeling condition, we derive the following upper bound for the radio number of the strong prismatic network with star containing odd cycle order. In advance, note that when m is odd the $\operatorname{diam}(G)=\frac{m-1}{2}$.

\begin{theorem}
Let $ G = S_n \boxtimes C_m $ denote the strong prismatic network with star. For any sub-star $ G'' \subset G $, we have
\begin{equation}
rn(G)  \leq \frac{m^2 + 2mn + 3m - 6n + 10}{2}.
\end{equation}
\end{theorem}

\begin{proof}
Let $u_1$ and $v_1$ be the centers of $S_n(t)$ and $S_n(t+\frac{m-1}{2})$. By Definition 2.7, $d(u_1,v_1)=\frac{m-1}{2}$. Setting $\varphi(v_1)=0$ and using $\operatorname{diam}(G)=\frac{m-1}{2}$ for odd $m$, the radio labeling condition gives:
\begin{equation}
\begin{aligned}
\varphi(u_1) &\leq \varphi(v_1) + {diam}(G) + 1 - d(u_1, v_1) \\
&= \frac{m-1}{2} + 1 - \frac{m-1}{2} = 1.
\end{aligned}
\end{equation}

When $ t=2 $, let $ v_1=(v_0, u_2) $ and $ u_1'=(v_0, u_{2+\frac{m-1}{2}}) $. Then $ d(v_1, u_1')=\frac{m-1}{2} $. By Equation (13) and Definition 2.7, we obtain:
\begin{equation}
\begin{split}
\varphi(u_1') &\leq \varphi(v_1) + \frac{m+1}{2} + 1 - \frac{m-1}{2} \\
&= \varphi(v_1) + 1.
\end{split}
\end{equation}

From Definition 2.7, $ G $ has $ \frac{m-3}{2} $ sub-star $ G''(t) $. Summing these $ \frac{m-3}{2} $ inequalities obtain:
\begin{equation}
\begin{aligned}
rn(G'') = \varphi(u_1(k)) &\leq \frac{m-3}{2} \times (m+2n-3) + 2 \times \frac{m-5}{2} \\
&= \frac{m^2 + 2mn + 3m - 6n + 10}{2}.
\end{aligned}
\end{equation}

The theorem is proven.
\end{proof}

\begin{theorem}
Let $ G = S_n \boxtimes C_m $ denote the strong prismatic network with star, then the radio labeling of $G$ is

\begin{equation}
rn(G)  \leq \frac{m^2 + 4mn + 5m - 8n + 18}{2}.
\end{equation}
\end{theorem}

\begin{proof}
Define $ c_1 = (v_0, u_1) \in S_n(1) $ as the central vertex.
Let $ \varphi(v_1) \geq 0 $. For the leaf $ v_m = (v_k, u_m) $ of $ S_n(m) $, by the definition of star, $ d(c_1, v_m) = 1 $. Substituting into the equation (13), we obtain the label of $ v_m $ as:
\begin{equation}
\varphi(v_m) \leq \varphi(v_1) + \frac{m-1}{2} + 1 - 1 = \varphi(v_1) + \frac{m-1}{2}. 
\end{equation}
\vspace{-3mm}

Let $ v_{\frac{m+1}{2}} $ be a leaf of $ S_n\left(\frac{m+1}{2}\right) $. By the definition of the radio labeling and the definition of star, we obtain Therefore, the radio number of the vertex $v_{\frac{m+1}{2}}$ is:
\begin{equation}
\begin{aligned}
\varphi(v_{\frac{m+1}{2}}) &\leq \varphi(v_m) + \frac{m-1}{2} + 1 - 1 \\
&= \varphi(v_m) + 1.
\end{aligned}
\end{equation}
\vspace{-1mm}

For the central vertex of $ S_n\left(\frac{m+1}{2}\right) $, through Equations (33) and (34), its radio labeling:

\begin{equation}
\varphi((v_0, u_{\frac{m+1}{2}})) = \frac{1}{2} (2mn + m - 2n + 8). 
\end{equation}
\vspace{-2mm}

Since $ G = (G') \cup (G'') $, combining Theorem 3.3 we have:
\begin{equation}
\begin{aligned}
rn (G'') &\leq \frac{m^2 + 2mn + 3m - 6n + 10}{2} + \frac{2mn + 2m - 2n + 8}{2} \\
&= \frac{m^2 + 4mn + 5m - 8n + 18}{2}.
\end{aligned}
\end{equation}
\vspace{-3mm}

The theorem is proven.
\end{proof}

Having established the lower bound for the radio number of the odd-cycle strong prismatic network with star, we now prove the corresponding upper bound in Theorem 3.5.

\begin{theorem}
Let $ G = S_n \boxtimes C_m$ denote the strong prismatic network with star, then the radio labeling of $G$ is
\begin{equation}
rn(G) \geq \frac{m^2 + 4mn + 5m - 8n + 18}{2}.
\end{equation}
\end{theorem}

\begin{proof}
The analysis of the three path types in the characteristic path $P^{\prime}(t)$ allows us to prove the lower bound on vertex labels under the radio labeling constraint, which will be discussed under the following cases.

\noindent\hspace*{\parindent}\textbf{Case 1:} Consider the first special path $ P_1'(t) $. 
Let $ P_1'(t) = (v_1, u_1) \xrightarrow{\frac{m+1}{2}} (v_3, u_{\frac{m+1}{2}}) \xrightarrow{\frac{m+3}{2}} (v_2, u_m) $.  
For $v_1(1) = (v_1, u_1)$ with $\varphi(v_1(1)) \ge 0$, the distance in $G$ satisfies:  
\begin{equation}
\begin{aligned}
\varphi((v_2, u_m)) &\geq \varphi((v_1, u_1)) + \frac{m-1}{2} - 2 \\
&= \varphi((v_1, u_1)) + \frac{m-3}{2}.
\end{aligned}
\end{equation}

Similarly, by the star and strong product structure, the distance $ d((v_2, u_m), (v_3, u_{\frac{m+1}{2}})) = \frac{m-1}{2} $, thus we obtain:

\begin{equation}
\begin{aligned}
\varphi((v_3, u_{\frac{m+1}{2}})) &\geq \varphi((v_2, u_m)) + \frac{m-1}{2} + 1 - \frac{m-1}{2} \\
&= \varphi((v_2, u_m)) + 1.
\end{aligned}
\end{equation}

\noindent\hspace*{\parindent} \textbf{Case 2:} Consider the second special path $ P_2'(t) $.
Let $ P_2'(t) = (v_2, u_1) \xrightarrow{\frac{m+1}{2}} (v_0, u_{\frac{m+1}{2}}) \xrightarrow{\frac{m+3}{2}} (v_3, u_m) $.  
Set $v_1\left(\frac{m+1}{2}\right) = (v_0, u_{\frac{m+1}{2}})$ and $v_2(1) = (v_2, u_1)$ with $\varphi(v_2(1)) \ge 0$. The distance in $G$ satisfies $d((v_2, u_1), (v_3, u_m)) = 2$. Thus we obtain:
\begin{equation}
\begin{aligned}
\varphi((v_3, u_m)) &\geq \varphi((v_2, u_1)) + \frac{m-1}{2} + 1 - 2 \\
&= \varphi((v_2, u_1)) + \frac{m-3}{2}.
\end{aligned}
\end{equation}

Similarly, $ d((v_3, u_m), (v_0, u_{\frac{m+1}{2}})) = \frac{m-1}{2} $. Substitute into the radio labeling definition:
\begin{equation}
\begin{aligned}
\varphi((v_0, u_{\frac{m+1}{2}})) &\geq \varphi((v_3, u_m)) + \frac{m-1}{2} + 1 - \frac{m-1}{2} \\
&= \varphi((v_3, u_m)) + 1.
\end{aligned}
\end{equation}

\noindent\hspace*{\parindent} \textbf{Case 3:} Consider the third special path $ P_3'(t) $.  
Let $ P_3'(t) = (v_a, u_1) \xrightarrow{\frac{m+3}{2}} (v_b, u_{\frac{m+1}{2}}) \xrightarrow{\frac{m+3}{2}} (v_c, u_m) $,
where $a, b, c$ are distinct indices satisfying $4 \le a, b, c \le n$. The structure of $G$ gives $d((v_a, u_1), (v_c, u_m)) = 2$. Substituting this into (13) leads to:
\begin{equation}
\varphi((v_c, u_m)) \geq \varphi((v_a, u_1)) + \frac{m-3}{2}.
\end{equation}

Following the same approach as above, $ d((v_c, u_m), (v_b, u_{\frac{m+1}{2}})) = \frac{m-1}{2} $. Substitute into the radio labeling definition:
\begin{equation}
\varphi((v_b, u_{\frac{m+1}{2}})) \geq \varphi((v_c, u_m)) + 1.
\end{equation}

It is further derived that:
\begin{equation}
\begin{aligned}
\varphi((v_b, u_{\frac{m+1}{2}})) &\geq \varphi((v_a, u_1)) + \frac{m-3}{2} + 1 \\
&= \varphi((v_a, u_1)) + \frac{m-1}{2}.
\end{aligned}
\end{equation}

For the central vertex contribution of $ G' = S_n(1) \cup S_n(\frac{m+1}{2}) \cup S_n(m) $, by the radio labeling definition and $ {diam}(G) = \frac{m-1}{2} $:
\begin{equation}
rn(G') \geq \varphi(V_m) + {diam}(G) + 1 - \frac{m-1}{2},
\end{equation}

\noindent with $ \varphi(v_m) $ as the maximum label of $ S_n(m) $, it can be obtained:
\begin{equation}
rn(G') \geq \frac{2mn + 2m - 2n + 8}{2}.
\end{equation}

Combining with Theorem 3.3, the total lower bound of $ rn(G) $ is:
\begin{equation}
\begin{aligned}
rn(G) = rn(G') + rn(G'') &\geq \frac{2mn + 2m - 2n + 8}{2} + \frac{m^2 + 2mn + 3m - 6n + 10}{2} \\
&= \frac{m^2 + 4mn + 5m - 8n + 18}{2}.
\end{aligned}
\end{equation}

This completes the proof of the theorem.
\end{proof}

\subsection{Optimal radio labeling of the strong prismatic network with star}

The bounds for the radio labeling of the strong prismatic network with star have been established in the preceding analysis. To determine the optimal radio labeling, it now suffices to examine the case where equality holds. The proof will address this condition through a case analysis based on cycle parity.
\begin{theorem}
Let $ G = S_n \boxtimes C_m$ denote the strong prismatic network with star, then the radio labeling of $G$ is:
\begin{equation}
rn(G)=
\begin{cases}
\dfrac{m^{2}+m(m-1)+(n-1)(m-2)}{2}, & \text{if $m$ is even and $m\ge 4$},\\[8pt]
\dfrac{m^{2}+4mn+5m-8n+18}{2},      & \text{if $m$ is odd and $m\ge 5$}.
\end{cases}
\end{equation}
\end{theorem}

Depending on the parity of the cycle, we analyze the following cases respectively.

\begin{proof}
According to Lemma 1.4 and Definition 1.5, the label of the central vertex and the label of the leaves satisfy the following condition for any pair of vertices:
\begin{equation}
|\varphi(u) - \varphi(v)| \ge {diam}(G) + 1 -d_G(u, v),
\end{equation}
\hspace*{\parindent}\textbf{Case 1:} $m$ is even.

Based on Definition~2.3 and Theorem~3.1, we have:  
\begin{equation}\label{eq:vertex-labels}
\varphi(c_j)=(j-1)\times\frac{m}{2}, \qquad 
\varphi(v_j(k))=\varphi(v_j)+\frac{m}{2}+(k-1)\times\frac{m-2}{2}.
\end{equation} 

For the central vertex $a=c_j$ and a leaf vertex $b=v_j(k)$ of the same substar, the star structure implies $d(a,b)=1$. Consequently, condition (53) applied to this pair gives:
\begin{equation}
\varphi(b) - \varphi(a) \geq \frac{m}{2} + 1 - 1 = \frac{m}{2}.
\end{equation}

However, in Theorem 3.2, we have already obtained:
\begin{equation}
\varphi(v_j(k)) \geq \varphi(c_j) + \frac{m}{2} + (k-1) \times \frac{m-2}{2}, 
\end{equation}

\noindent which is consistent with the radio labeling constraint.
Let $ a = v_j(l) $ and $ b = v_j(k) $. By the definition of star, $ d(a,b) = 2 $. Based on Equation (53), we have:
\begin{equation}
|\varphi(b) - \varphi(a)| \geq \frac{m-2}{2}.
\end{equation}

From the definition of radio labeling, we obtain the following formal inequality for leaves within the same substar:
\begin{equation}\label{eq:leaf-label-diff}
|\varphi(v_j(l)) - \varphi(v_j(k))| \geq |l - k| \times \frac{m-2}{2}.
\end{equation}

Next, we verify that this scheme satisfies the constraint of radio labeling that $|\phi(u) - \phi(v)| \ge \operatorname{diam}(G) + 1 - d_G(u,v)$ holds for all pairs of vertices $u, v$. This can be accomplished by following the analysis:

\noindent\hspace*{\parindent} \textbf{Subcase 1.1: }Both points are central vertices.
Let $ a = c_i $ and $ b = c_j $. Then, $ d(a, b) = t $. Vertices $ a $ and $ b $ satisfy $ |\varphi(b) - \varphi(a)| \geq \frac{m}{2} + 1 - t $. We further obtain:
\begin{equation}
|\varphi(c_i) - \varphi(c_j)| \geq |i-j| \times \frac{m}{2} = t \times \frac{m}{2}. 
\end{equation}

It is easy to verify that this is the radio labeling constraint when $t \geq 1$.

\noindent\hspace*{\parindent}\textbf{Subcase 1.2: }One of the two points is a central vertex.
Let $a = c_i$ and $b = v_j(k)$ be two vertices from distinct substars, with distance $d(a, b) = t$. By the radio labeling condition, their labels therefore satisfy:
\begin{equation}\label{eq:radio-condition-distinct}
|\varphi(b) - \varphi(a)| \geq \frac{m}{2} + 1 - t.
\end{equation}

Similarly, applying Theorem~3.2 gives:
\begin{equation}
|\varphi(c_j) - \varphi(v_j(k))| \geq |\varphi(c_i) - \varphi(c_j)| \geq t \times \frac{m}{2}. 
\end{equation}

\noindent\hspace*{\parindent}\textbf{Subcase 1.3: }Both points are leaves.
Based on the definition of radio labeling and the distance relation between vertices, we formalize the label constraints for two leaf vertices $a = v_i(l)$ and $b = v_j(k)$ as follows.

Let $a = v_i(l)$ and $b = v_j(k)$ be two leaf vertices with distance $d(a, b) = t$. By the radio labeling condition, their labels satisfy:
\begin{equation}
|\varphi(b) - \varphi(a)| \geq \frac{m}{2} + 1 - t. \label{eq:leaf-dist}
\end{equation}

Furthermore, if $v_i$ and $v_j$ denote the central vertices of their respective substars and $d(v_i, v_j) = t$, then the lower bound provided by Theorem~3.1 gives:
\begin{equation}
|\varphi(v_i) - \varphi(v_j)| \geq t \times \frac{m}{2}. \label{eq:central-dist}
\end{equation}

This formal step clarifies the lower bounds for the label differences under different scenarios, laying the groundwork for verifying the overall labeling scheme.

In $ G $, the minimum radio labeling $ \varphi(v_1) = 0 $, and the maximum label appears at the last leaf of the final sub-star:
\begin{equation}
\begin{aligned}
\varphi(V_m(n)) &= \varphi(v_m) + \frac{m}{2} + (n-1) \times \frac{m-2}{2} \\
&= \frac{m^2 + m(m-1) + (n-1)(m-2)}{2}.
\end{aligned} 
\end{equation}

Having thoroughly analyzed the case where $m$ is even, we now proceed to examine case 2.

\noindent\hspace*{\parindent}\textbf{Case 2:} $m$ is odd.

The central vertex of sub-star $ S_n(j) $ is defined as $ c_j = (v_0, u_j) $, with label assignment:
\begin{equation}
\varphi(c_j) = (j-1) \times \frac{m-1}{2},
\end{equation}
define the leaves of $ S_n(j) $ as $ v_j(k) = (v_k, u_j) $ ( $ 1 \leq k \leq n $ ), with label assignment:
\begin{equation}
\varphi(v_j(k)) = \varphi(v_j) + \frac{m-1}{2} + (k-1).
\end{equation}

We verified the radio labeling constraint by considering the following subcase:

\noindent\hspace*{\parindent}\textbf{Subcase 2.1: }Vertices in the same sub-star.
Let $ a = v_j $ and $ b = v_j(k) $. By the star definition, $ d(a, b) = 1 $. According to the radio labeling definition:
  \begin{equation}
  |\varphi(b) - \varphi(a)| = \frac{m-1}{2} + (k-1). 
  \end{equation}

Since the required minimum difference for radio labeling is $ {diam}(G) + 1 - d(a,b) = \frac{m-1}{2} $, and $ \frac{m-1}{2} + (k-1) \geq \frac{m-1}{2} $, this is the required constraint.

Let $ a = v_j(l) $ and $ b = v_j(k) $. By the star definition, $ d(a, b) = 2$. According to the radio labeling definition:
  \begin{equation}
  |\varphi(b) - \varphi(a)| = |k - l|.
  \end{equation}

The required minimum difference is: 
\begin{equation}
\text{diam}(G) + 1 - d(a,b) = \frac{m-1}{2} + 1 - 2 = \frac{m-3}{2}.
\end{equation}

\noindent\hspace*{\parindent}\textbf{Subcase 2.2:} Vertices from different sub-star.
Let $ a = c_i $ and $ b = v_j(k) $. By the strong prismatic network with star structure, $ d(a, b) = 1 $. According to the radio labeling definition:
  \begin{equation}
  |\varphi(b) - \varphi(a)| = \frac{m-1}{2} \times |i - j|,
  \end{equation}
which meets the required minimum difference, thus satisfying this constraint.

For all vertex pairs $a=v_i(l)$, $b=v_j(k)$ and other combinations, the label difference obtained from the assignment rule meets or exceeds the required minimum, thereby satisfying the radio labeling constraint.

Let $ a = v_1 $ and $ b = v_2 $. By the star definition, $ d(a, b) = 1 $ and $ \varphi(b) = \frac{m-1}{2} $. From the radio labeling definition:
\begin{equation}
|\varphi(b) - \varphi(a)| = \frac{m-1}{2} + 1 - 1 = \frac{m-1}{2}.
\end{equation}
  
Let the central vertex of $ S_n(m) $ be $ v_0 $ and the largest leaf be $ v_m $. The sub-star $ S_n(m) $ contains $ m $ layers of leaves. By Definition 2.9, $ d = \frac{m-1}{2} $. 
Thus, the radio labeling of $v_m$ is obtained:
\begin{equation}
\begin{aligned}
\varphi(v_m) &= \frac{1}{2} [ m^2 + 2n(m-1) - 5m - 4(n-1) ] \\
&= \frac{1}{2}  (m^2 + 2mn - 6n - 5m + 4 ).
\end{aligned}
\end{equation}
 
Let the central vertex of $ S_n(\frac{m+1}{2} ) $ be $ c_0 $ and the largest leaf be $ v_{\frac{m+1}{2}} $. sub-star $ S_n( \frac{m+1}{2} ) $ contains $ \frac{m+1}{2} $ layers. The radio labeling of the vertex $v_{\frac{m+1}{2}}$ is:
\begin{align}
\varphi(v_{\frac{m+1}{2}}) &= \frac{2mn + 10m - 2n + 14}{2}. 
\end{align}
 
For $m \geq 4$, the maximum radio labeling depends on both $\varphi(v_{\frac{m+1}{2}})$ and $\varphi(v_m)$, allowing us to determine the radio labeling $rn(G)$ of the graph.
\begin{equation}
\begin{aligned}
rn(G) &= \varphi(v_{\frac{m+1}{2}}) + \varphi(v_m) \\
&= \frac{2mn + 10m - 2n + 14}{2} + \frac{1}{2}(m^2 + 2mn - 6n - 5m + 4) \\
&= \frac{m^2 + 4mn + 5m - 8n + 18}{2}.
\end{aligned}
\end{equation}

By analyzing the properties of the sub-star, we ultimately obtain the radio labeling for $G$ when $m$ is odd:
$rn(G) = \frac{m^2 + 4mn + 5m - 8n + 18}{2}.$
The theorem is proven.
\end{proof}

We have established tight bounds for the radio number of the strong prismatic network with star and constructed an optimal labeling, thus proving the main theorem.

Let $S_n$ denote $n$-star and $C_m$ denote $m$-cycle. Then the optimal radio labeling of $G$ is:
\begin{equation}
rn(G)=
\begin{cases}
\dfrac{m^{2}+m(m-1)+(n-1)(m-2)}{2}, & \text{if $m$ is even and $m\ge 4$},\\[8pt]
\dfrac{m^{2}+4mn+5m-8n+18}{2},      & \text{if $m$ is odd and $m\ge 5$}.
\end{cases}
\end{equation}

Thus, we have completed the proof of the radio labeling for the strong prismatic network with star. To better understand Theorem 3.6 presented in this paper, we provide simple Example 1 and Example 2.

\begin{figure}[htbp]
    \centering
    \begin{minipage}[b]{0.4\textwidth}
        \centering
        \includegraphics[width=\linewidth]{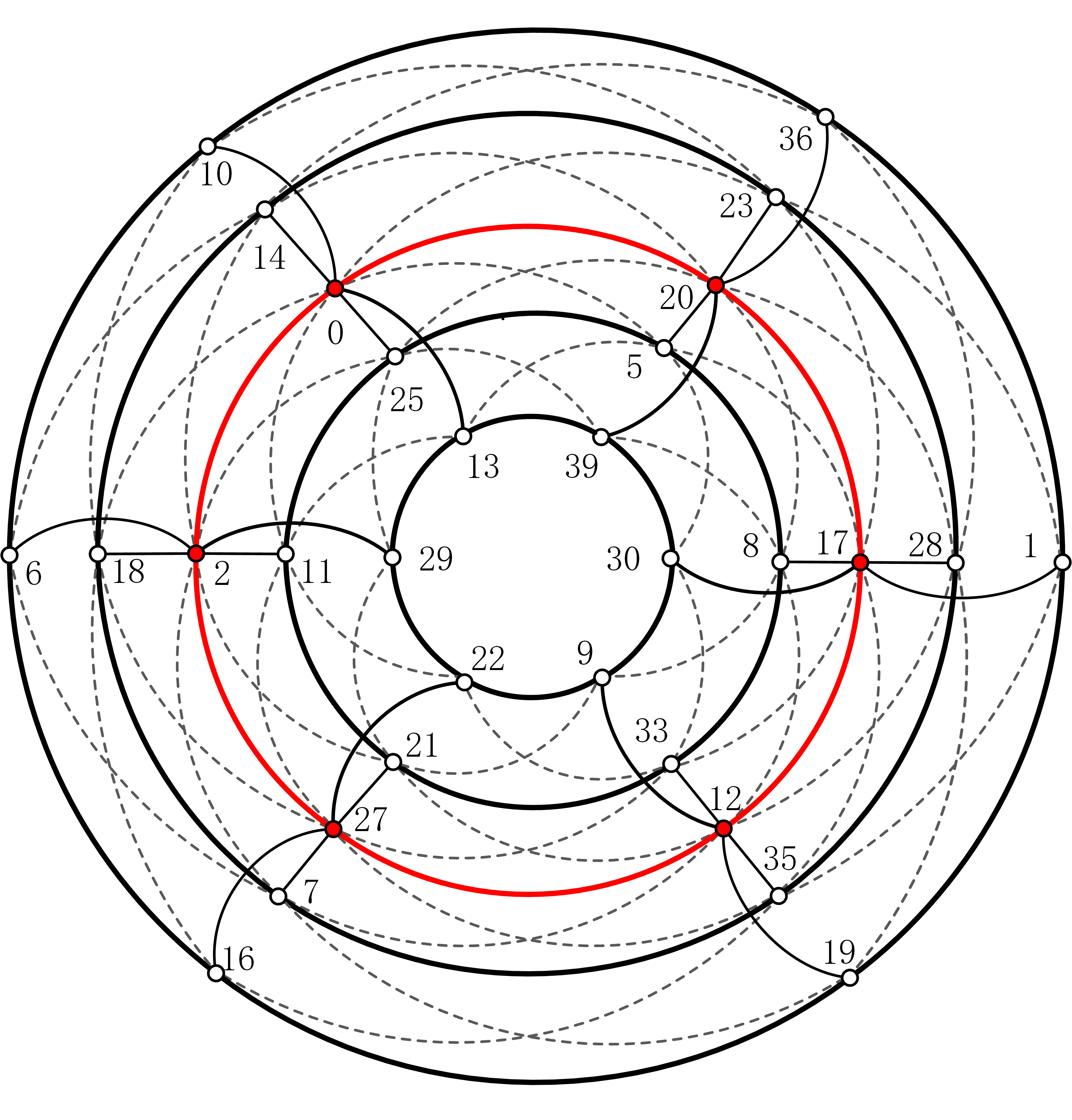}
        \caption{Radio labeling of strong prismatic network with star when m = 6.}
        \label{fig:example1}
    \end{minipage}
    \hfill
    \begin{minipage}[b]{0.4\textwidth}
        \centering
        \includegraphics[width=\linewidth]{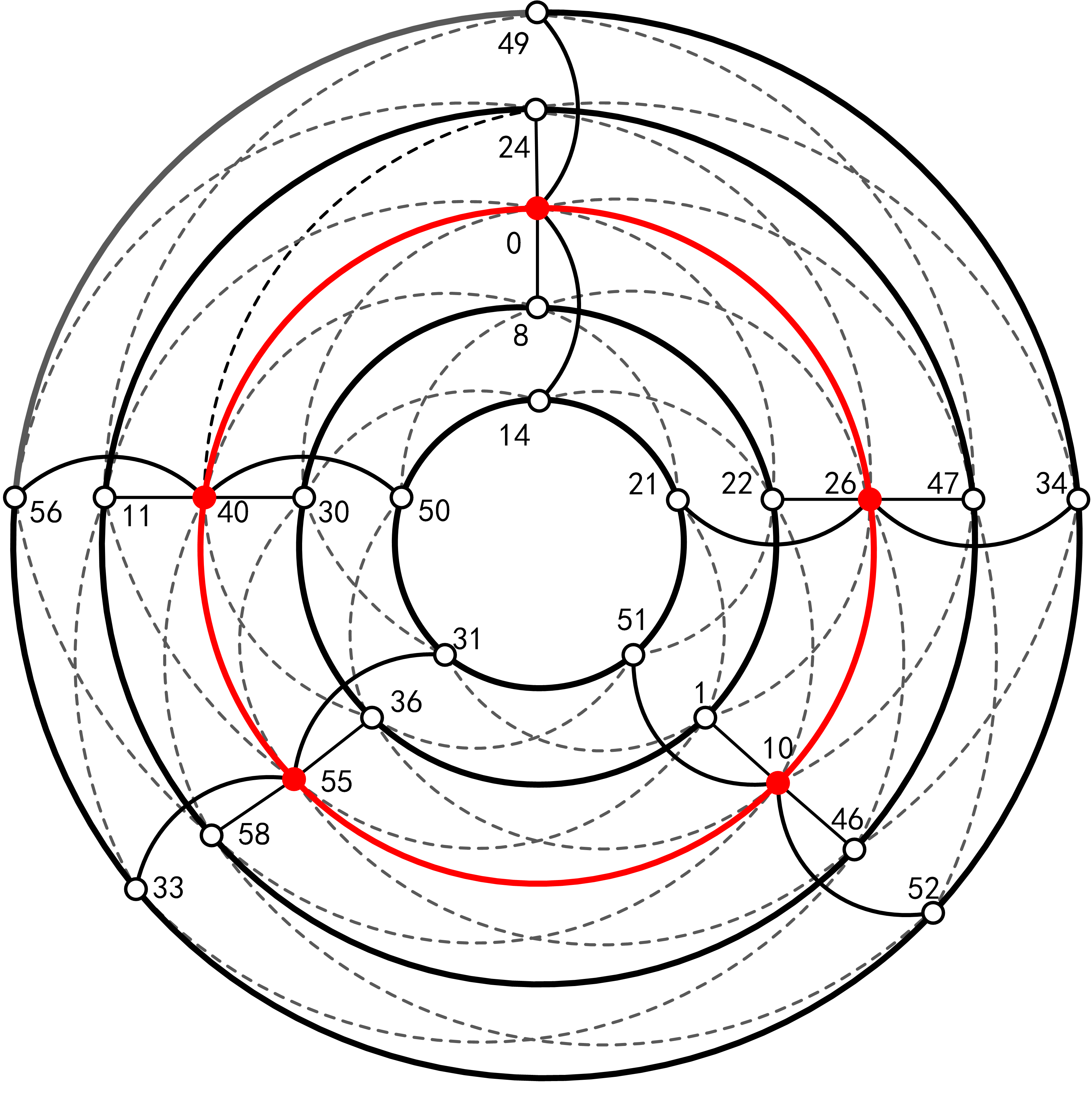}
        \caption{Radio labeling of strong prismatic network with star when m = 5.}
        \label{fig:example2}
    \end{minipage}
\end{figure}

\textbf{Example 1}: Suppose 30 communication sites need to communicate with each other. To reduce unnecessary interference during communication, a channel assignment scheme is adopted. This scheme combines a 4-order star with a 6-order cycle through the strong product operation, forming a topological structure. As shown in Figure \ref{fig:example1}, this scheme achieves optimal channel assignment, enabling interference-free communication between sites. The radio labeling for this network is 39.

\textbf{Example 2}: Suppose 25 communication sites need to communicate with each other. To reduce unnecessary interference during communication, a channel assignment scheme is adopted. This scheme combines a 4-order star with a 5-order cycle through the strong product operation, forming a topological structure. This achieves an optimal channel assignment, enabling interference-free communication between sites, as shown in Figure \ref{fig:example2}. The radio labeling for this network is 58.

Since the calculation of vertex labels in the strong prismatic network with star becomes extremely cumbersome when the orders of the star and cycle are sufficiently large, we have designed the polynomial algebraic proof process of the main theorem into an algorithm and implemented it through MATLAB, with the detailed implementation provided in the appendix.

This algorithm supports parallel computation and can process multiple (n, m) parameter combinations simultaneously. It is suitable for scenarios that require the batch computation of different parameter combinations, significantly improving computational efficiency, especially for large-scale parameter ranges.

\begin{figure}[htbp]
    \centering
    \begin{minipage}[b]{0.45\textwidth}
        \centering        \includegraphics[width=\linewidth]{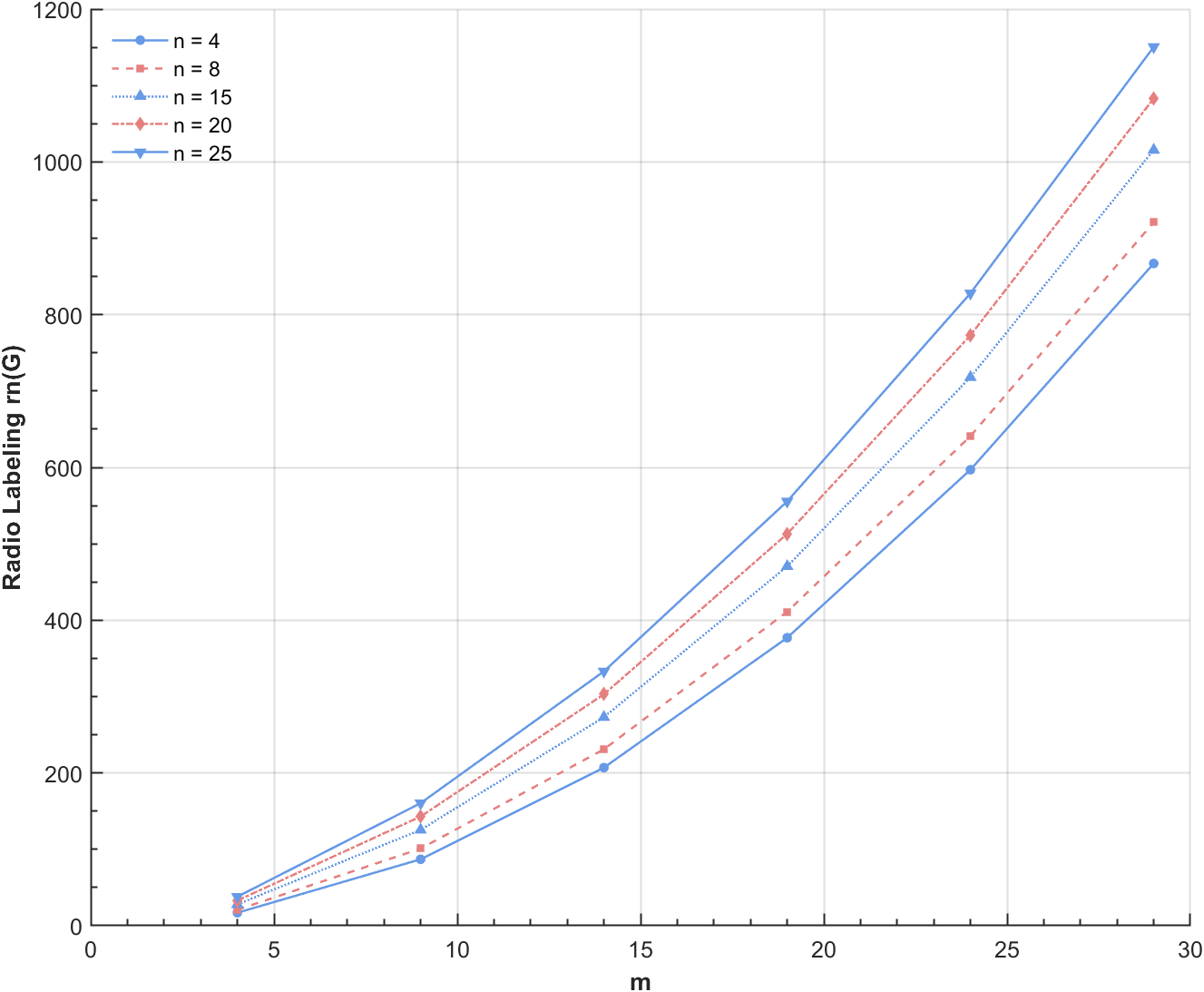}
        \caption{Radio labeling of strong prismatic network with star when  $m$  is even.}
        \label{fig:even}
    \end{minipage}
    \hfill
    \begin{minipage}[b]{0.45\textwidth}
        \centering
        \includegraphics[width=\linewidth]{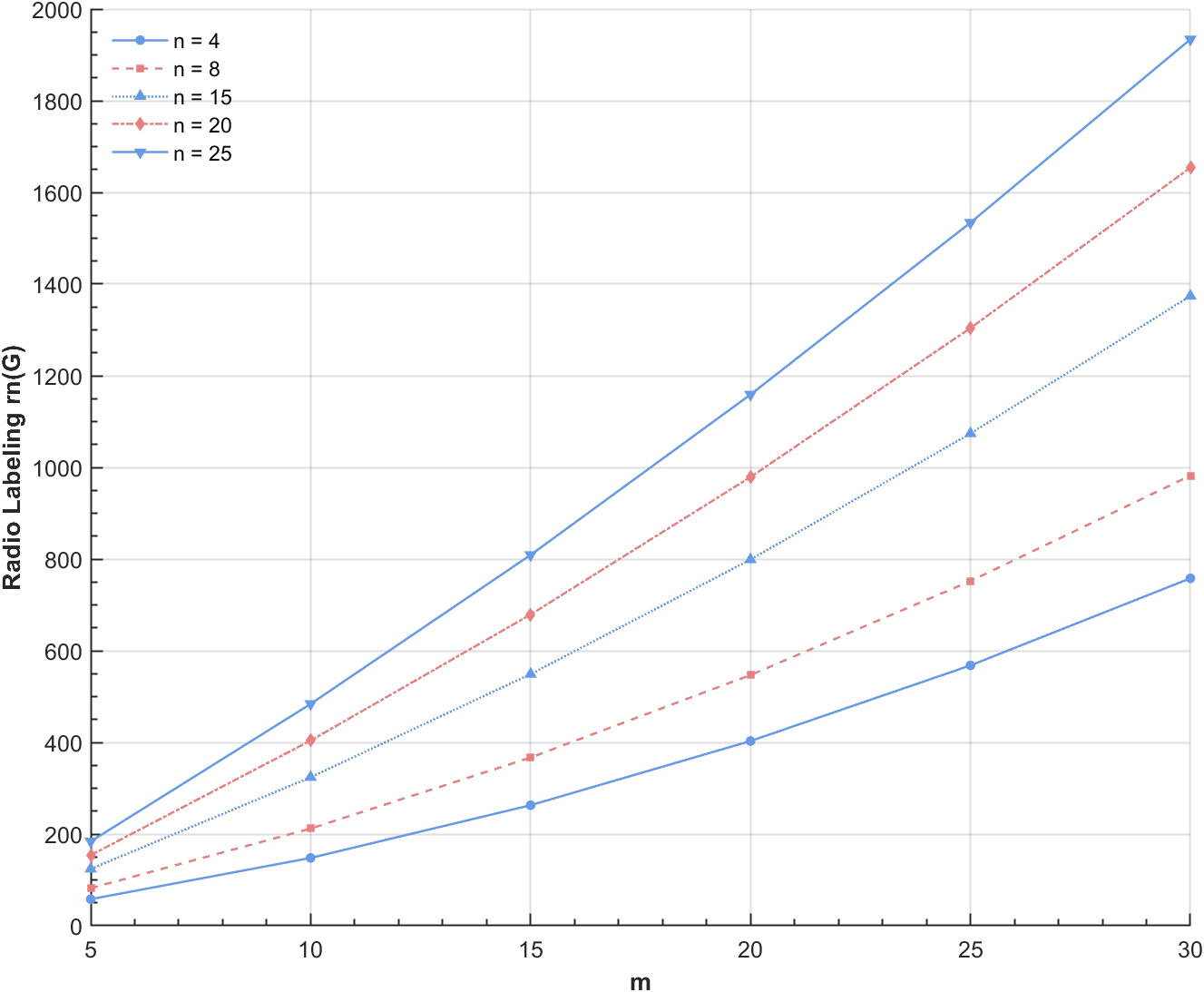}
        \caption{Radio labeling of strong prismatic network with star when  $m$  is odd.}
        \label{fig:odd}
    \end{minipage}
\end{figure}

Figure \ref{fig:even} and Figure \ref{fig:odd} illustrate the functional relationship between the optimal radio labeling $\mathrm{rn}(G)$ and the parameters $n$ and $m$, visually demonstrating the theoretical results of Theorem 3.6. Under identical scale parameters, the strong prismatic network with star achieves a smaller optimal radio labeling number when m is even, thus rendering it superior to the case where m is odd in both theoretical performance and practical applications.

\begin{figure}[!ht]
    \centering
    \includegraphics[width=0.75\linewidth]{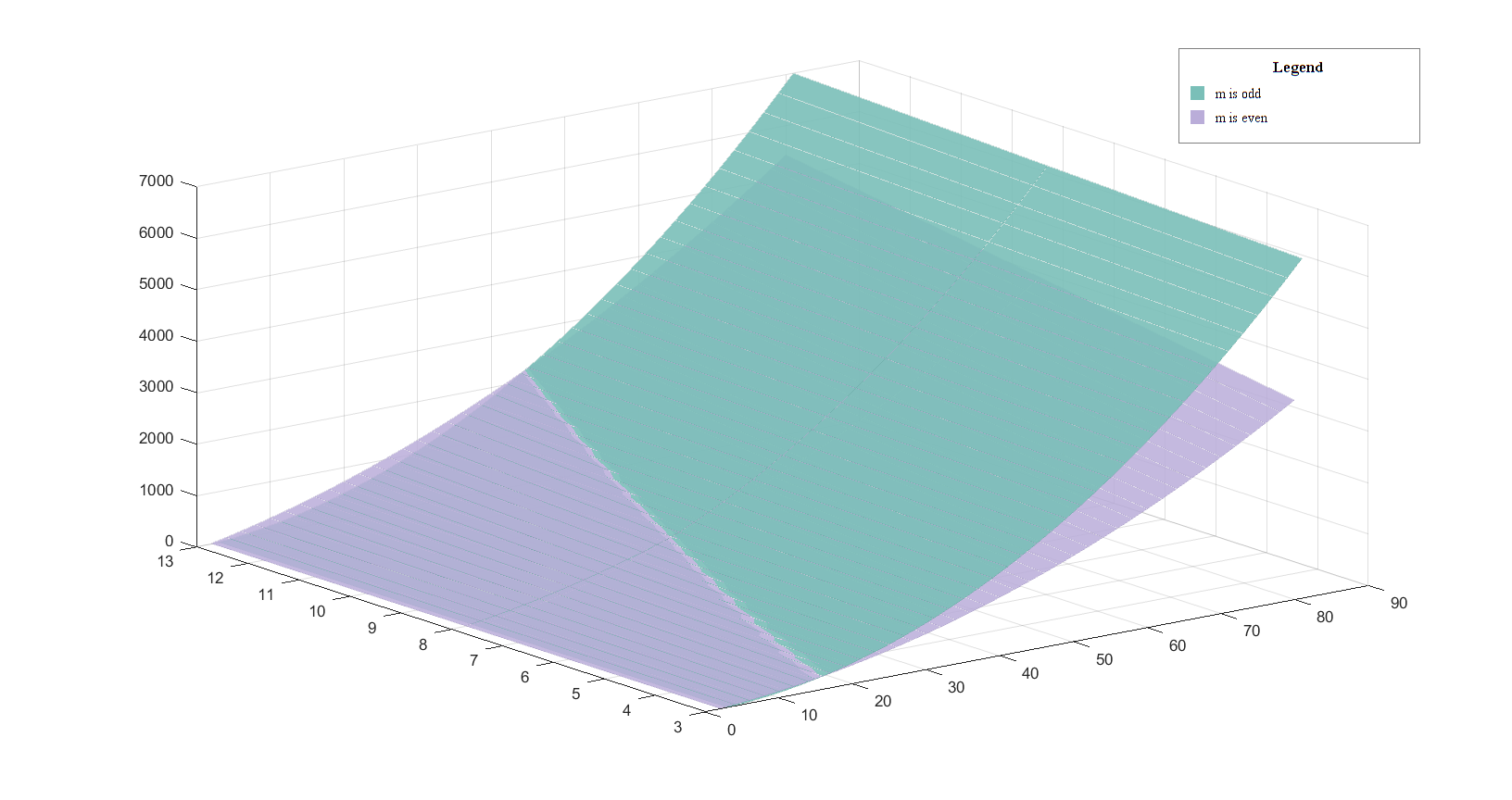}
    \caption{Visualization model of the strong prismatic network with star.}
    \label{fig:last}
\end{figure}
The visualization outcomes unambiguously reveal that within the hybrid configurations of strong prismatic network with star, the optimal radio labeling demonstrates a marked ascending progression with increasing values of both parameters $m$ and $n$. Furthermore, the parity of the cycle order manifests a pronounced and systematic impact on the behavior of $rn(G)$. Under sufficiently large values of $m$ and $n$, the $rn(G)$ values associated with even order cycle substantially surpass those corresponding to their odd order counterparts.

Analyzing the visualization model diagram reveals that when $m$ is even, the radio labeling of strong prismatic network with star exhibits significantly better performance. Therefore, in practical applications, it is recommended to use even-order cycles to construct such networks, thereby achieving more compact and efficient spectrum assignment.

\section{Conclusion}

This study focuses on strong prismatic network with star. By analyzing their structural properties, we have precisely computed the radio number ${rn}(G)$ for the topological structure of this specific family of graphs. and have constructed an optimal labeling scheme. The study has involved analyzing the parity of $m$ to determine the bounds of the radio labeling, and the functional dependence of ${rn}(G)$ on $n$ and $m$ was elucidated, supported by line charts and a visualization graph. These findings present a significant solvable case for the NP-hard radio labeling problem, with direct implications for improving communication efficiency, minimizing interference, and optimizing the utilization of wireless channels. Future research will investigate spectrum assignment in other strong prismatic network to enhance channel utilization.

\section*{Acknowledgment}

This work received partial support from the National Natural Science Foundation of China under Grant No. 11551002, as well as partial support from the Natural Science Foundation of Qinghai Province under Grant No. 2019-ZJ-7093.
\vspace{4mm}

{\bf \small Data Availability}: Our manuscript has no associated data.

\vspace{4mm}
{\bf \small Declaration of interests}: The authors affirm that there are no known competing financial interests or personal connections of theirs that could have influenced the research outlined in this paper.

\vspace{4mm}
{\bf \small  Author contributions}: Conceptualization and methodology, Feng Li and Liming Wang; formal analysis and investigation, Feng Li, Liming Wang, and Linlin Cui; resources, Feng Li; writing—original draft preparation, Feng Li and Liming Wang; writing—review and editing, Feng Li, Liming Wang, and Linlin Cui; supervision, project administration, and funding acquisition, Feng Li. All authors have read and approved the final manuscript.


\appendix
\section{Appendix}
\begin{breakablealgorithm}
\caption{Parallel Computation of Radio Labeling for strong prismatic network with star}
\label{alg:radioStarCycleParallel}

\renewcommand{\algorithmicrequire}{\textbf{Input:}}
\renewcommand{\algorithmicensure}{\textbf{Output:}}

\begin{algorithmic}[1]
\Require Parameter ranges: $n_{\text{min}}, n_{\text{max}}, m_{\text{min}}, m_{\text{max}}$, number of trials $t$
\Ensure Optimal radio labeling matrix $m(G)$, performance metrics, visualization data

\State \textbf{Initialization:}
\State Set random seed for reproducibility: $\text{seed} \leftarrow 42$
\State Open output file: $f_{\text{out}} \leftarrow \text{fopen}(\text{"radio\_labeling\_output.txt"})$
\State Print system information: MATLAB version, operating system, available memory
\State Initialize parallel pool with dynamic worker allocation

\State \textbf{Serial Computation Function:} $\text{radio\_labeling\_serial}(n, m)$
\State Validate input: $n \geq 4$ and integer, $m \geq 4$ and integer
\If{$m \bmod 2 = 0$ and $m \geq 4$}
    \State $rn(G) \leftarrow (m^2 + m(m-1) + (n-1)(m-2)) / 2$ 
\ElsIf{$m \bmod 2 = 1$ and $m \geq 5$}
    \State $rn(G) \leftarrow (m^2 + 4mn + 5m - 8n + 18) / 2$ 
\Else
    \State Return $\text{NaN}$ 
\EndIf

\State \textbf{Parallel Computation Function:} $\text{radio\_labeling\_parallel}(\mathbf{n}, \mathbf{m})$
\State Validate input vectors: $\mathbf{n}$ and $\mathbf{m}$ as vectors
\State $N \leftarrow \text{length}(\mathbf{n}), M \leftarrow \text{length}(\mathbf{m})$
\State Initialize result matrix: $\mathbf{R} \leftarrow \text{zeros}(N, M)$

\Statex \textbf{Conceptual parallel computation:} \Comment{Each iteration independent; suitable for parallelization}
\For{$i = 1$ to $N \times M$}
    \State Compute indices: $(row, col) \leftarrow \text{ind2sub}([N, M], i)$
    \State $n \leftarrow \mathbf{n}(row), m \leftarrow \mathbf{m}(col)$
    \If{$m \bmod 2 = 0$ and $m \geq 4$}
        \State $\mathbf{R}(i) \leftarrow (m^2 + m(m-1) + (n-1)(m-2)) / 2$
    \ElsIf{$m \bmod 2 = 1$ and $m \geq 5$}
        \State $\mathbf{R}(i) \leftarrow (m^2 + 4mn + 5m - 8n + 18) / 2$
    \Else
        \State $\mathbf{R}(i) \leftarrow \text{NaN}$
    \EndIf
\EndFor

\State Return $\mathbf{R}$

\State \textbf{Benchmark Testing:} $\text{benchmark\_radio\_labeling}(n_{\text{max}}, m_{\text{max}}, t)$
\State Generate test data: $\mathbf{n}_{\text{test}} \leftarrow \text{randi}([4, n_{\text{max}}], 1, n_{\text{max}}-3)$
\State Generate test data: $\mathbf{m}_{\text{test}} \leftarrow \text{randi}([4, m_{\text{max}}], 1, m_{\text{max}}-3)$
\State \textbf{Serial timing:} $t_{\text{serial}} \leftarrow \text{zeros}(1, t)$
\For{$k = 1$ to $t$}
    \State Start timer: $t_{\text{start}} \leftarrow \text{tic}()$
    \For{$i = 1$ to $\text{length}(\mathbf{n}_{\text{test}})$}
        \For{$j = 1$ to $\text{length}(\mathbf{m}_{\text{test}})$}
            \State $\text{radio\_labeling\_serial}(\mathbf{n}_{\text{test}}(i), \mathbf{m}_{\text{test}}(j))$
        \EndFor
    \EndFor
    \State $t_{\text{serial}}(k) \leftarrow \text{toc}(t_{\text{start}})$
\EndFor
\State \textbf{Parallel timing:} $t_{\text{parallel}} \leftarrow \text{zeros}(1, t)$
\For{$k = 1$ to $t$}
    \State Start timer: $t_{\text{start}} \leftarrow \text{tic}()$
    \State $\mathbf{R}_{\text{test}} \leftarrow \text{radio\_labeling\_parallel}(\mathbf{n}_{\text{test}}, \mathbf{m}_{\text{test}})$
    \State $t_{\text{parallel}}(k) \leftarrow \text{toc}(t_{\text{start}})$
\EndFor
\State Compute statistics: $\mu_{\text{serial}}, \sigma_{\text{serial}}, \mu_{\text{parallel}}, \sigma_{\text{parallel}}$
\State Compute speedup: $S \leftarrow \mu_{\text{serial}} / \mu_{\text{parallel}}$
\State Print benchmark results to console and output file

\State \textbf{Scaling Analysis:}
\State Define problem sizes: $\mathbf{P} \leftarrow [100, 500, 1000, 2000]$
\For{$p \in \mathbf{P}$}  
    \State $\mathbf{n}_{\text{scale}} \gets 4:\min(4+p, 500)$
    \State $\mathbf{m}_{\text{scale}} \gets 4:\min(4+p, 500)$
    \State Time serial computation for all $(n, m)$ pairs
    \State Time parallel computation using $\text{radio\_labeling\_parallel}$
\EndFor
\State Compute speedup ratios: $\mathbf{S}_{\text{scale}} \leftarrow \mathbf{t}_{\text{serial}} \oslash \mathbf{t}_{\text{parallel}}$
\State Generate scaling plots: execution time vs. problem size, speedup vs. problem size

\State \textbf{Example Calculations:}
\State Compute individual examples: $(n,m) \in \{(4,4), (4,5), \ldots, (20,13)\}$
\State Batch computation: $n \in [4,8], m \in [4,9]$
\State Sequence calculation: $n=4$ fixed, $m \in [4,15]$
\State Pattern analysis: $n \in [4,10], m \in [4,10]$
\State Output all results in formatted tables to console and file

\State \textbf{Visualization:}
\State Generate surface plot: $rn(G) = f(n,m)$
\State Generate heatmap: color-coded $rn(G)$ values
\State Mark special points: $(4,4), (10,10), (20,20), (4,20), (20,4)$
\State Add contour lines and annotations

\State \textbf{Parity Analysis:}
\State Fix $n = 10$
\State $\mathbf{m}_{\text{even}} \leftarrow [4,6,8,\ldots,20]$
\State $\mathbf{m}_{\text{odd}} \leftarrow [5,7,9,\ldots,21]$
\For{$i = 1$ to $\min(\text{length}(\mathbf{m}_{\text{even}}), \text{length}(\mathbf{m}_{\text{odd}}))$}
    \State Compute $rn(G)$ for $m_{\text{even}}(i)$ and $m_{\text{odd}}(i)$
\EndFor
\State Compute growth rates:
    \State $g_{\text{even}} \leftarrow (rn(G)_{\text{even,end}} - rn(G)_{\text{even,start}}) / ({rn}_{\text{even,end}} - {rn}_{\text{even,start}})$
    \State $g_{\text{odd}} \leftarrow (rn(G)_{\text{odd,end}} - rn(G)_{\text{odd,start}}) / ({rn}_{\text{odd,end}} - {rn}_{\text{odd,start}})$
\State Compute ratio: $r \leftarrow g_{\text{odd}} / g_{\text{even}}$
\State Generate parity comparison plot and growth rate visualization

\State \textbf{Performance Visualization:}
\State Generate bar chart: serial vs. parallel mean execution times with error bars
\State Generate speedup chart: parallel speedup factor with baseline
\State Generate trial-by-trial performance plot with mean lines

\State \textbf{Cleanup:}
\State Close parallel pool
\State Close output file: $\text{fclose}(f_{\text{out}})$
\State Display completion message

\Ensure Complete set of $rn(G)$ values, performance metrics, visualization data, output file
\end{algorithmic}
\end{breakablealgorithm}

The core computation of this algorithm has a time complexity of $O(1)$, relying only on basic arithmetic operations in a closed-form formula; the serial time complexity for batch parameter scanning is $O(NM)$, where $N$ and $M$ are the numbers of values for $n$ and $m$, respectively. The data parallelism achieved through MATLAB parfor significantly reduces the actual runtime, completing 774,400 calculations in only 0.82 seconds with a peak memory usage of 4.8 MB.

To investigate the influence of the parity of the cycle length $m$ on the radio number of the star, we fix the parameter $n=10$ and calculate the corresponding $rn(G)$ for even and odd values of $m$, as shown in Table \ref{tab:parity_analysis_complete}. It can be observed that regardless of whether $m$ is even or odd, $rn(G)$ strictly increases monotonically with $m$. Moreover, for comparable values of $m$, the $rn(G)$ corresponding to odd $m$ is significantly higher than that for even $m$. Further linear regression analysis of the growth rates in Table \ref{tab:growth_rate_analysis} shows an average growth rate of 28.00 for even $m$ and 35.50 for odd $m$, with a ratio of approximately 1.27. This indicates that the growth rate of $rn(G)$ under odd cycle lengths is about 1.27 times that under even cycle lengths. These findings are consistent with the theoretical derivation: the closed-form formula for odd $m$ contains higher-order cross terms, which make $rn(G)$ more sensitive to variations in $m$.

\begin{table}[htbp]
\centering
\begin{minipage}[t]{0.48\textwidth}
\centering
\caption{Comparing even vs odd $m$ for fixed $n=10$}
\label{tab:parity_analysis_complete}
\begin{tabular}{cccc}
\toprule
\textbf{Even $m$} & \textbf{$rn(G)$} & \textbf{Odd $m$} & \textbf{$rn(G)$} \\
\midrule
4 & 23.00 & 5 & 94.00 \\
6 & 51.00 & 7 & 151.00 \\
8 & 87.00 & 9 & 212.00 \\
10 & 131.00 & 11 & 277.00 \\
12 & 183.00 & 13 & 346.00 \\
14 & 243.00 & 15 & 419.00 \\
16 & 311.00 & 17 & 496.00 \\
18 & 387.00 & 19 & 577.00 \\
20 & 471.00 & 21 & 662.00 \\
\bottomrule
\end{tabular}
\end{minipage}
\hfill
\begin{minipage}[t]{0.48\textwidth}
\centering
\caption{Growth rate analysis for even and odd $m$ values ($n=10$)}
\label{tab:growth_rate_analysis}
\begin{tabular}{lc}
\toprule
\textbf{Parameter} & \textbf{Value} \\
\midrule
Growth rate for even $m$ & 28.00 \\
Growth rate for odd $m$  & 35.50 \\
Odd/Even growth ratio & 1.27 \\
\bottomrule
\end{tabular}
\end{minipage}
\end{table}

\begin{figure}[!ht]
    \centering
    \includegraphics[width=0.95\linewidth]{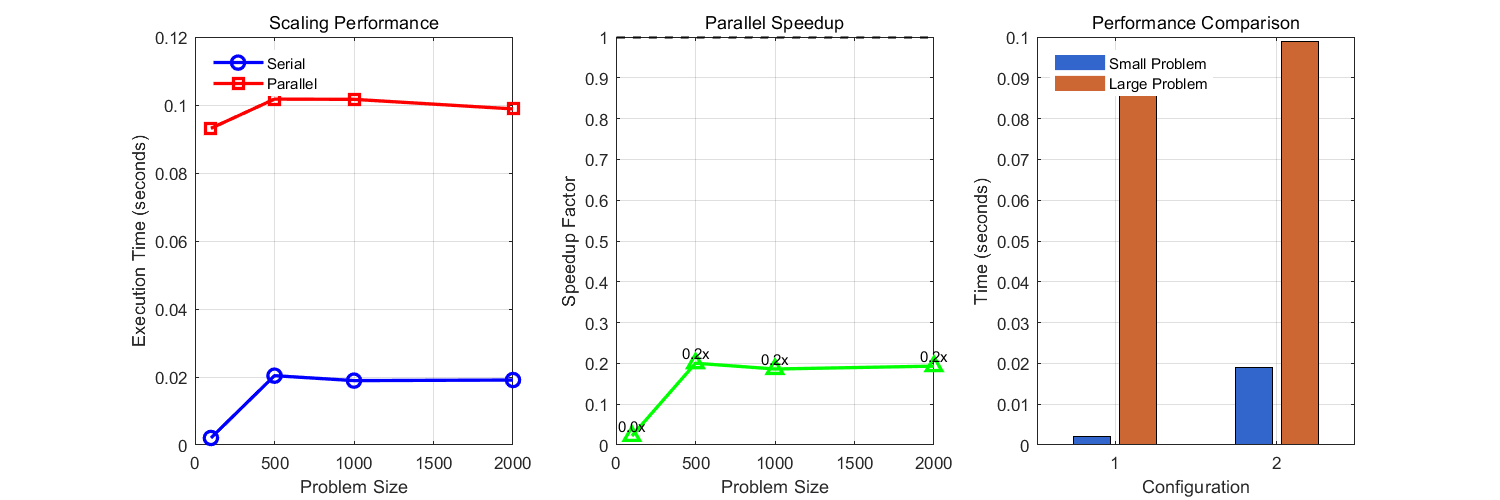}
    \caption{Performance analysis of serial and parallel Implementations.}
    \label{fig:placeholder7}
\end{figure}

As shown in Figure \ref{fig:placeholder7}, the algorithm not only possesses a solid theoretical foundation but also demonstrates notable computational performance. In a single‑core environment, it achieves relatively fast solving speeds for medium‑scale problems. When a parallelization strategy is introduced, it maintains efficient processing even for large‑scale parameter tasks, with computation times consistently kept at the millisecond level. The ability to quickly solve small‑scale problems and stably handle large‑scale ones highlights the feasibility of the closed‑form formulation, thus providing efficient and reliable computational support for investigating the radio labeling properties of strong prismatic network with star.

\newpage
\parbox[t]{\linewidth}{
\noindent\parpic{\includegraphics[height=1.2in,width=1in,clip,keepaspectratio]{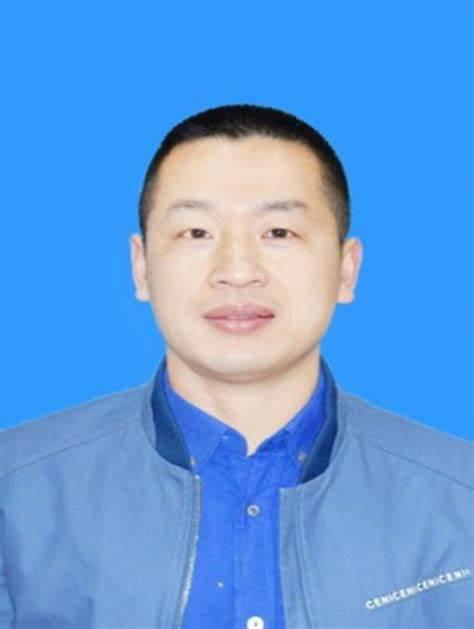}}
\noindent {\bf Feng Li} was born in Anhui province, China. He is a tenured professor teaching at the Qinghai normal university. He supervises PhD students, doctoral supervision. He received his Ph.D. in 2013 from the school of mathematics and statistics, Xi'an Jiaotong University. His research interests include graph theory, wireless communication, optimization theory and algorithms of interconnection network, and machine learning. He has been presided and participated more than 15 projects including the National Natural Science Foundation of China (NSFC), Qinghai Natural Science Foundation of China (NSFC) and Ministry of Education's Program. He has published over 118 academic research papers, he has obtained 11 European and Chinese patents for invention in combinatorial network design, and he has obtained more than 127 software copyrights in graph theory and combinatorial network.}

\vspace{10mm}

\parbox[t]{\linewidth}{
\noindent\parpic{\includegraphics[height=1.2in,width=1in,clip,keepaspectratio]{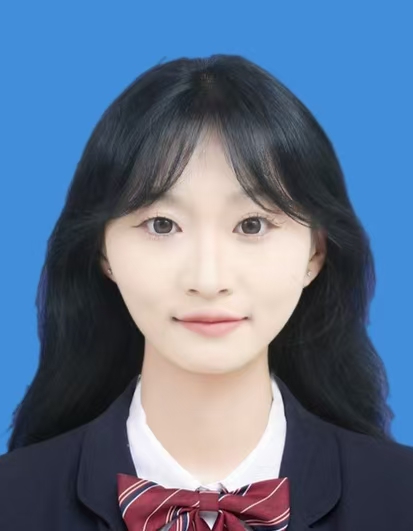}}
\noindent {\bf Liming Wang} was born in Gansu province, China, in 2002. She is currently pursuing the M.Sc. in the School of Computer Science, Qinghai Normal University, China. She received her bachelor's degree in 2024 from the School of Information Engineering, Lanzhou Industry and Commerce College, Lanzhou, China. Her research interests include graph theory, analysis and design of reliable combinatorial network.}

\vspace{20mm}

\parbox[t]{\linewidth}{
\noindent\parpic{\includegraphics[height=1.2in,width=1in,clip,keepaspectratio]{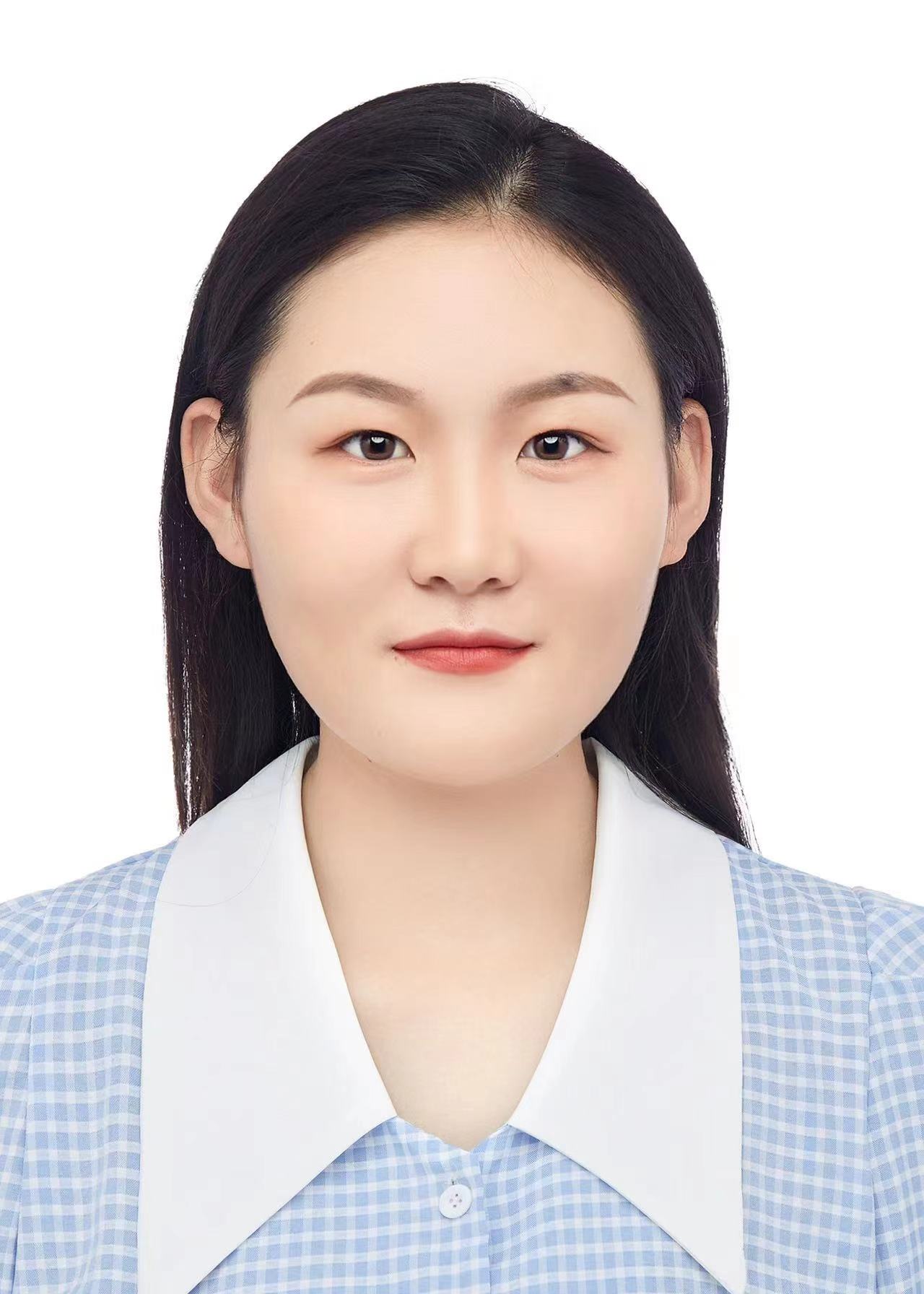}}
\noindent {\bf Linlin Cui} was born in Henan province, China, in 1998. She is currently pursuing the Ph.D. in the School of Computer Science, Qinghai Normal University, China. She received her master's degree in 2024 from School of Computer Science, Qinghai Normal University. Her research interests include graph theory, analysis and design of reliable combinatorial network.}
\end{CJK} 
\end{document}